\documentclass[twocolumn]{emulateapj}

\usepackage{amsmath,natbib,appendix}
\usepackage{longtable}
\usepackage[autoplay]{animate}
\usepackage{graphicx}
\usepackage{xcolor}
\slugcomment{Submitted to ApJ}

\shorttitle{HYDROGEN MOLECULES IN THE DARK AGES HALOS}

\shortauthors{NOVOSYADLYJ ET AL.}

\def\x{$\times$}

\def\h2o{H$_2$O}
\def\so2{SO$_2$}
\def\13co{$^{13}$CO}
\def\nh3{NH$_{3}$}
\def\hco+{HCO$^{+}$}
\def\water18{H$_2^{18}$O}

\begin{document}
\title{HYDROGEN MOLECULES IN THE DARK AGES HALOS: \\ THERMAL EMISSION VS. RESONANT SCATTERING}

\author{B. Novosyadlyj$^{1,2}$, V. Shulga$^{1,3}$,  Yu. Kulinich$^2$, W. Han$^{1}$}
 
\affil{$^1$College of Physics and International Center of Future Science of Jilin University, Qianjin Street 2699, Changchun, 130012, P.R.China, \\
$^2$Astronomical Observatory of Ivan Franko National University of Lviv, Kyryla i Methodia str., 8, Lviv, 79005, Ukraine,\\
$^3$Institute of Radio Astronomy of NASU, 4 Mystetstv str., 61002 Kharkiv, Ukraine}

\begin{abstract}
{The emission from dark ages halos in the lines of transitions between lowest rotational levels of hydrogen and hydrogen deuteride molecules is analyzed.   It is assumed molecules to be excited by CMB and collisions with hydrogen atoms. The physical parameters of halos and number density of molecules are precalculated in assumption that halos are homogeneous top-hat  spheres formed from the cosmological density perturbations in the four-component Universe with post-Planck cosmological parameters. The differential brightness temperatures and differential spectral fluxes in the rotational lines of H$_2$-HD molecules are computed for two phenomena: thermal luminescence  and resonant scattering of CMB radiation. The results show that expected maximal values of differential brightness temperature of warm  halos ($T_K\sim$200-800 K) are at the level of nanokelvins, are comparable for both phenomena, and are below sensitivity of modern sub-millimeter radio telescopes. For hot halos ($T_K\sim$2000-5000 K) the thermal emission of H2-ortho molecules dominates and the differential brightness temperatures are predicted to be of a few microkelvins  at the frequencies 300-600 GHz, that could be detectable with telescopes of a new generation. }
 
\end{abstract}
 
\section{Introduction}

Formation of the first luminous objects of the Universe is important topic of current cosmology. It is the generally accepted paradigm that they are formed from the initial small matter density and velocity perturbations generated at the early stages of the evolution of the Universe. The most interesting stages of their early evolution are hidden in the Dark Ages, period of time between the last scattering of the cosmic microwave background (CMB) at $z\approx1000$ and reionization of the intergalactic medium by early luminous objects at $z\approx10$ \citep{Planck2018a}. The theories of generation and evolution of cosmological perturbations successfully describe the observed anisotropy of CMB and the large-scale structure of the Universe  that make it possible to  evaluate the basic  parameters of the cosmological model of the Universe and the ratio of the densities of its main components such as baryonic matter, dark matter and dark energy, with sufficient accuracy. The lowest scales in these theories, which are probed by observations, are dozens of Megaparsecs in current astronomical units that are essentially larger than expected ones for the first luminous objects, stars, globular clusters or dwarf galaxies. The numerical simulations of the large-scale structure formation at these scales are very complicated and ambiguous because the non-linear dynamical effects in the multicomponent medium, gas-dynamics, molecular chemistry and cooling/heating processes become important or even key. The considerable ambiguity of the predictions of theoretical simulations at these scales is due also to the uncertainty of the physical nature of dark matter and dark energy. This supports the importance of analyzing the possibilities of detection of signals from Dark Ages both in the line of 21 cm of atomic hydrogen and in the lines of the first molecules.

The theoretical investigations of the possible signal from the Dark Ages in the hyperfine structure line 21 cm of atomic hydrogen are intensive and already have long history (see, for example, reviews \cite{Barkana2001,Fan2006,Furlanetto2006a,Pritchard2012}). It was predicted that an expected signal at $\sim100$ MHz can be at the level of hundreds - dozens of millikelvins of brightness temperature. It seems that the long-term efforts of several scientific groups to detect such signal from the Dark Ages gave finally the first results: the team of the Experiment to Detect the Global EoR  Signature\footnote{https://www.haystack.mit.edu/ast/arrays/Edges/} (EDGES) has announced the registration of absorption line 21 cm of atomic hydrogen with the brightness temperature about 0.5 K at $z=$15-20 \citep{Bowman2018}. Unfortunately the signal in the molecular lines is expected rather weaker but it can be more informative since it can say about molecular chemistry and physical conditions of excitation of lines. The upper limits for such signal at the level of few millikelvins have been obtained in a few observations by IRAM 30 m telescope \citep{deBernardis1993}, RATAN-600 \citep{Gosachinskij2002} and the Odin satellite \citep{Persson2010}.

The molecular line signals from the Dark Ages can be observed in the emission or in absorption on the background of the CMB radiation, that is defined by competition of excitation/de-excitation of levels by the CMB radiation and the local sources of radiation or collisions with particles. So, the local physics conditions in the baryonic matter of Dark Ages may cause the secondary anisotropy of CMB. This mechanism is known as thermal emission/absorption and it is like to photo- or thermoluminescence in the solids \citep{Chen1997}. The emission in the rotational lines of H$_2$ and HD molecules from the forming Population III objects and protogalaxies in Dark Ages have been studied by \citet{Kamaya2002,Kamaya2003,Omukai2003,Mizusawa2005} and other authors. The marginal possibility of their detection by current and future facilities and their importance have been discussed by them too. The gas clouds or halos which have large peculiar velocity along the line of sight are another source of secondary anisotropy: the resonant scattering of CMB quanta by molecules in the moving halos may cause increasing or decreasing of the brightness temperature of the radiation passing through the halo depending on the direction of peculiar velocity relative to the observer. It was predicted by \cite{Dubrovich1977} and possibility of its observation were studied in \citet{Maoli1994,Maoli1996,Dubrovich2008,Nunez2006,Persson2010} and other papers. 

The gas in the halos, which have been virialized at $10\le z\le 50$, is over-densed and re-heated \citep{Novosyadlyj2018}: the range of matter density is $\sim10^{-24}-10^{-22}$ g/cm$^3$, the range of gas temperature heated adiabatically is $\sim50-800$ K, heated by shocks in the processes of violent relaxation to virial temperature is $\sim10^4-10^5$ K. Hence, the dark ages halos can be a detectable source of thermal emission in the molecular lines of the most abundant molecules H$_2$ and HD. In the paper we estimate the brightness temperature of molecular lowest rotational energy levels emission from dark ages halos under the condition of radiative and collisional excitation.

The outline of this paper is as follows. In section 2 we briefly describe the models of halos, their physical characteristic and chemical composition. In section 3 we describe the computations of the rates of collisional excitations of the lowest five rotational levels of hydrogen molecules by atomic hydrogen and estimate the critical number density of perturbers. The methods of computations of populations and excitation temperatures of these levels are presented in section 4. In section 5 we present the results of computations of opacity, differential brightness temperatures and differential spectral fluxes caused by thermal collisions of hydrogen molecules with hydrogen atoms. In section 6 we compute for the same halos and lines the differential brightness temperatures and fluxes caused by the resonant scattering of the CMB radiation. Conclusions are in section 7.

The table with data for all halos analyzed in the paper, tables with coefficients of  interpolations, tables with the main results and animate figures with evolution of opacities and differential brightness temperatures for halos of different mass are presented in the Appendix of the on-line version.

\section{Models of dark ages halos}

\begin{table*}[!ht]
\begin{center}
\caption{Physical values and chemical composition of halos virialized at different $z_v$: $M$ is the total mass, $C_k$ is the amplitude of initial curvature perturbation (seed of halo), $z_v$ is the redshift of virialization, $\rho_{m}$ is the matter density virialized halo, $T_K$ is kinetic temperature of baryonic gas, $n_{HI}$ is the number density of neutral hydrogen atoms, $n_p,\,n_e$ are the number densities of protons and electrons at $z=z_v/10$, $n_{H_2}$ and $n_{HD}$ are the number densities of molecules H$_2$ and HD, $r_h$ is the radius of halo in comoving coordinates, $\theta_h$ is the angular radius of geometrically limited halo.}
\begin{tabular} {cccccccccccc}
\hline
\hline
   \noalign{\smallskip}
$M_h$&$k$&$C_k$&$z_v$&$\rho_{m}$&$T_K$&$n_{HI}$&$n_p\approx n_e$&$n_{H_2}$&$n_{HD}$&$r_h$&$\theta_h$\\
 \noalign{\smallskip} 
[M$_{\odot}$]&[Mpc$^{-1}$]& & &[g/cm$^3$]&[K]&[cm$^{-3}$]&[10$^{-6}$cm$^{-3}$]&[10$^{-6}$cm$^{-3}$]&[10$^{-9}$cm$^{-3}$]&[kpc]&[arcsec]\\
 \noalign{\smallskip} 
\hline
   \noalign{\smallskip} 
$5.3\cdot10^9$&5&$3.0\cdot10^{-4}$&30.41&$1.52\cdot10^{-23}$&402.1&1.14&106.2/3.8&14.39&2.51&1.78&1.03\\ 
    \noalign{\smallskip}
              & &$2.5\cdot10^{-4}$&25.15&$8.79\cdot10^{-24}$&298.9&0.66& 66.1/4.0&5.99&2.07&2.14&1.05\\ 
    \noalign{\smallskip}
              & &$2.0\cdot10^{-4}$&19.90&$4.49\cdot10^{-24}$&206.3&0.34& 36.7/4.0&2.15&1.51&2.68&1.09\\  
    \noalign{\smallskip}
              & &$1.5\cdot10^{-4}$&14.65&$1.89\cdot10^{-24}$&124.3&0.14& 17.0/4.4&0.63&0.52&3.60&1.15\\  
    \noalign{\smallskip}
              & &$1.0\cdot10^{-4}$& 9.41&$5.55\cdot10^{-25}$& 59.8&0.04& 5.6/5.0&0.13&0.08&5.38&1.26\\    
  \hline
\end{tabular}
\label{Tab1}
\end{center}
\end{table*}

We set the physical conditions and chemistry of the halos by modeling the evolution of individual spherical perturbations in the four-component Universe (cold dark matter, baryon matter, dark energy, and thermal relict radiation) starting from the linear stage at the early epoch, through the quasi-linear stage, turnaround point and collapse up to virialized state \citep{Novosyadlyj2016,Novosyadlyj2018}. All physical values and chemical composition of a halo with mass $M_h=5.3\cdot10^9$ M$_\odot$, which are necessary for computation of the excitations and brightness temperatures in the molecular rotational lines, are presented in Table \ref{Tab1}. The data for halos of other mass, $6.6\cdot10^8$, $8.3\cdot10^7$, $1.0\cdot10^7$ and $1.3\cdot10^6$M$_\odot$, are presented in Table \ref{Tab1A} in Appendix of the on-line version.

All computations in the paper are performed for consistent values of the main parameters of the cosmological model, namely, the Hubble constant $H_0=70$ km/s/Mpc, the mean density of baryonic matter in the units of critical one $\Omega_b=0.05$, the mean density of dark matter $\Omega_m=0.25$, the mean density of dark energy $\Omega_{de}=0.7$, its equation of state parameter $w_{de}=-0.9$, the effective sound speed $c_s=1$ (in units of speed of light). The models with  dark energy with $w=const$ and cold dark matter (CDM) often are noted as wCDM. The dark matter can be also warm dark matter (WDM) with mass of particles larger than a few keV.  

We suppose that halo is spherical and homogeneous (top-hat) with values of matter density, kinetic temperature and the number density of species obtained from the computations of its formation. The mass of each halo $M_h$ in the solar mass, its radius in comoving coordinates $r_h$ [kpc] and the wave number $k$ [Mpc$^{-1}$] of initial perturbation from which halo is formed, are connected by relations
\begin{equation}
\frac{M_h}{M_{\odot}}=1159\Delta_v(1+z_v)^3\Omega_mh^2r_h^3=4.5\cdot10^{12}\Omega_mh^2k^{-3},\nonumber
\end{equation}
where $z_v$ is the redshift of halo virialization, $\Delta_v\equiv\rho_m(z_v)/\bar{\rho}_m(z_v)$, $\Omega_m\equiv\bar{\rho}_m(0)/\rho_{cr}(0)$, 
$h\equiv H_0/100\rm{km/s/Mpc}$. They are presented in Table \ref{Tab1} and \ref{Tab1A}. The angular sizes of analyzed halos are in the range $\sim0.06-1.25$ arcsec (last columns of tables \ref{Tab1} and \ref{Tab1A}). In the computations we assume $\Delta_v=178$. 

The values of matter density, kinetic temperature of baryon component and radius of halo, presented in Table \ref{Tab1}, do not change after virialization. The number density of neutral hydrogen atoms ($n_{HI}$) is practically unchanged, while the number densities of molecules, protons ($n_p$) and electrons ($n_e$) are monotonically changed since molecular reactions continue. Therefore, we present their values in Table \ref{Tab1} and \ref{Tab1A} at $z_v$ and $z=10$, in the computations we use the model values at any $z$.

\section{Rates of excitation/de-excitation of rotational levels of H$_2$ and HD}

\begin{table}[!ht]
\begin{center}
\caption{The values of spontaneous transitions allowed by quantum selection rules, frequencies and energy for the lowest five rotational energy levels of hydrogen molecules H$_2$ and hydrogen deuteride molecules HD.}
\begin{tabular} {ccccccc}
\hline
\hline
   \noalign{\smallskip}
Species&Transitions\footnote{Here and below $u$ means upper level, $l$ means lower level}&A$_{ul}$&  Frequency $\nu_{ul}$& E$_u$\\
 \noalign{\smallskip}
       &$J_u$\,--\,$J_l$ &[s$^{-1}$] & [GHz]      & [K] \\
  \noalign{\smallskip}
\hline
   \noalign{\smallskip} 
H$_2$&2\,--\,0&2.94\x10$^{-11}$	&10\,621& 510   \\
     &3\,--\,1&4.76\x10$^{-10}$	&17\,594&1\,015  \\
     &4\,--\,2&2.75\x10$^{-9}$ 	&24\,410&1\,681\\
     &5\,--\,3&9.83\x10$^{-9}$	&31\,011&2\,503 \\
   \noalign{\smallskip}      
HD   &1\,--\,0&5.32\x10$^{-8}$ 	&2\,675 &128   \\
     &2\,--\,1&5.05\x10$^{-7}$	&5\,332 &384    \\
     &3\,--\,2&1.80\x10$^{-6}$	&7\,952 &766    \\
     &4\,--\,3&4.31\x10$^{-6}$	&10\,518&1\,271 \\
     &5\,--\,4&8.35\x10$^{-6}$	&13\,015&1\,895 \\
 \noalign{\smallskip}
  \hline
\end{tabular}
\label{Tab2}
\end{center}
\end{table}

In this paper we analyze the possible line emission of molecules H$_2$ and HD caused by spontaneous transitions between the lowest rotational levels with quantum numbers $J=0,\,1,\,...5$ in the dark ages halos before reionization at $z\ge10$. They can be excited by quanta of CMB, collisions of these molecules with atoms and molecules, and slight light background of first stars at the end of Dark Ages. The rotational quantum numbers of allowed transitions $J_u-J_l$, Einstein coefficients $A_{ul}$,frequencies $\nu_{ul}$ and energy of levels\footnote{http://www.cv.nrao.edu/php/splat/advanced.php}  $E_u$ of molecules H$_2$ and HD are presented in Table \ref{Tab2}.  

\begin{figure*}
\includegraphics[width=0.49\textwidth]{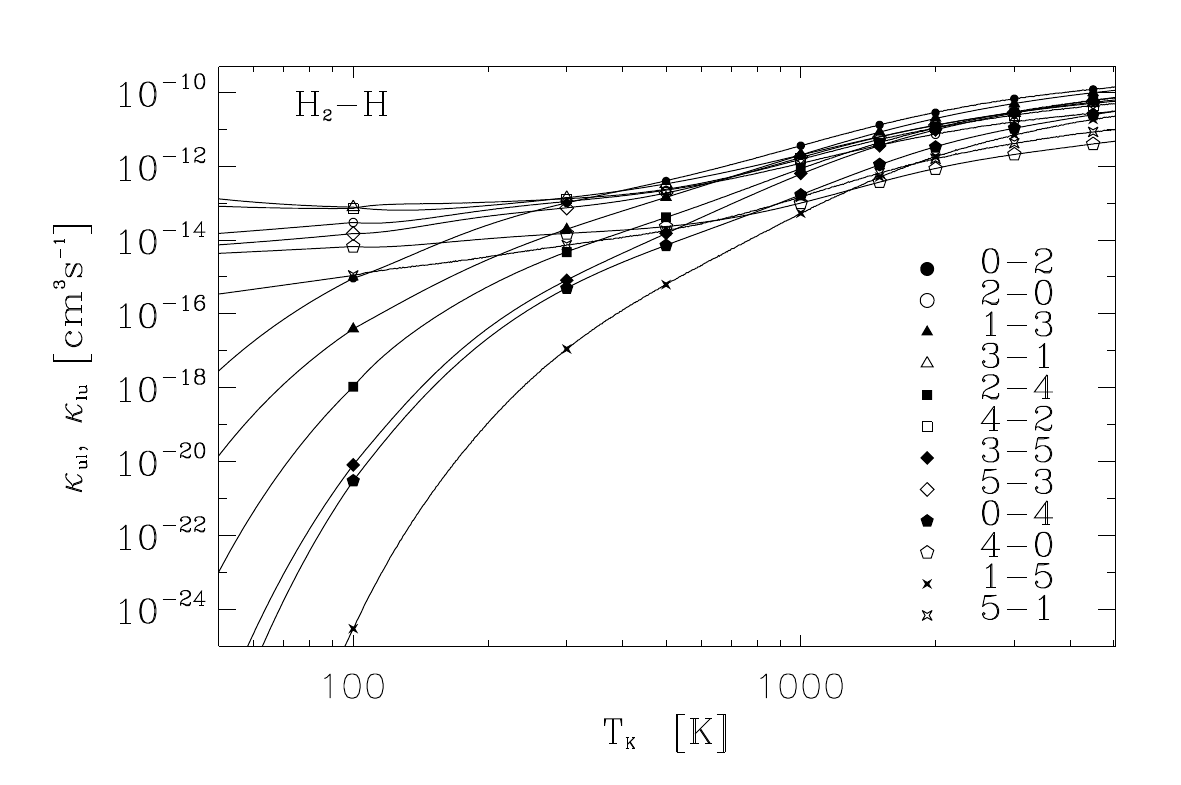}
\includegraphics[width=0.49\textwidth]{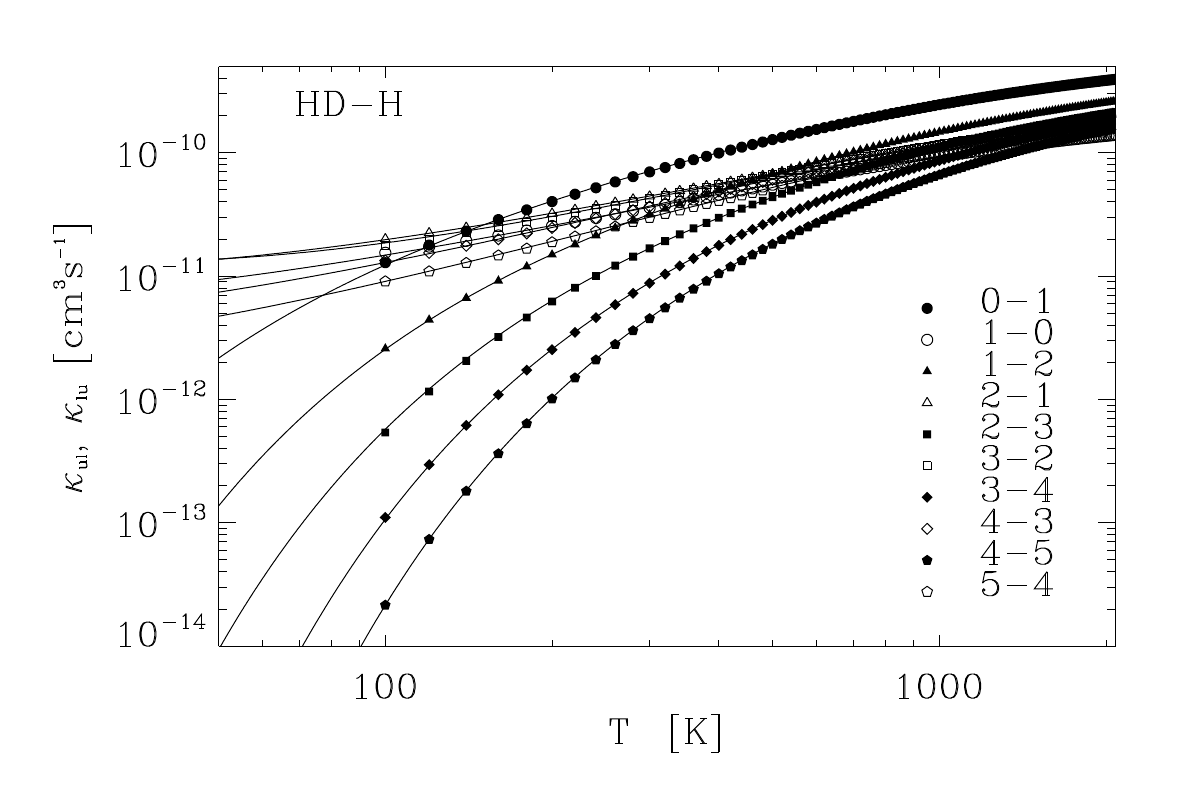}
\caption{Rate coefficients for collisional excitation ($\kappa_{lu}$, filled symbols) and de-excitation ($\kappa_{ul}$, open symbols) of molecules H$_2$ and HD by H.}
\label{VAMDCkappa}
\end{figure*}

The collisional excitations are crucially important since they make a possibility to separate the luminous of halos from the CMB background, by other words, to detect them. 
The bases of the theory of collisional excitation of lowest rotational levels of molecules, which are important for observational radioastronomy, have been initiated in the papers \cite{Purcell1952,Takayanagi1960,Takayanagi1963,Field1966,Rogers1968,Goss1968}. The computations of excitation/de-excitation rate of lowest ro-vibrational levels of molecules H$_2$ by collisions with hydrogen atoms have been done by \citet{Flower1997,Flower1998,Wrathmall2007,Lique2012,Lique2015}. The data by \citet{Flower1997,Flower1998,Wrathmall2007} contain the rate coefficients of collision transitions between ro-vibrational levels with $\Delta J=\pm2,\,\pm4,\, ...$ up to 26th level for both spin isomers ortho- and para-hydrogen molecule. They are available at the Ro-Vibrational Collisional Excitation
Database of VAMDC consortium\footnote{https://basecol.vamdc.eu/} as well as the analytical fits of their dependences on temperature which are enough accurate for the temperature range 100-6000 K. We use these fits for the de-excitation rate coefficients of para-/ortho-H$_2$ by atomic hydrogen for $T_K\ge 100$ K, for the lower temperatures we use the extrapolation by second-order polynom on the base of nearest three datapoints. The coefficients for fits and extrapolations of collisional deexcitation $\kappa_{ul}$ of lowest rotational levels of molecules para-H$_2$ ($J=0,\,2,\,4$) and ortho-H$_2$ ($J=1,\,3,\,5$) used here are presented in Appendix of the on-line version in Tables \ref{TabH2deexcfit} and \ref{TabH2lowTcdexc} accordingly. The collisional excitation rate coefficients $\kappa_{lu}$ are computed using the well known relation $\kappa_{lu}=\kappa_{ul}g_u/g_l/\exp{(h\nu_{ul}/kT_K)}$. The dependences of coefficients of collisional rate excitation/de-excitations of three lowest levels of para-/ortho-H$_2$ molecules by atomic hydrogen on temperature are shown in the left panel of Figure \ref{VAMDCkappa}. We take into account the transition ortho$\leftrightarrows$para using ratio $n_{ortho}/n_{para}$ computed by \cite{Flower2000} for the temperature range $1-10^4$ K. We also estimate the impact of new collisional coefficients by \citet{Lique2015} on the excitation temperatures of rotational transitions of ortho-/para-H$_2$ and brightness temperatures in the most intensive lines.  

The computations of excitation/de-excitation rate coefficients of the lowest ro-vibrational levels of hydrogen deuteride molecules HD by collisions with hydrogen atoms have been done by \citet{Flower1999,Roueff1999}.  
The complete results for the most transitions with $\Delta J=1,\,2,\,... $ for levels  with $J=0,\,...\,9$ and $v=0,\,1,\,2$ are available at the database of VAMDC consortium for the temperature range 100-2080 K with step $\Delta T_K=20$ K. They are well approximated by cubic polynoms in the log-log scale that is shown in the right panel of Figure \ref{VAMDCkappa}. The best-fit coefficients are presented in Table \ref{TabHDTcdexc} in the Appendix of the on-line version. For the temperatures lower 100 K we use the extrapolated values issued by this fit. 

The rates of collisional excitation/de-excitation of $u-l$ levels of molecules H$_2$ and HD (noted as X) are computed as follows
$$C^X_{ul}=\kappa^X_{ul}n_H, \quad C^X_{lu}=\kappa^X_{lu}n_H.$$

\begin{figure*}
\includegraphics[width=1.0\textwidth]{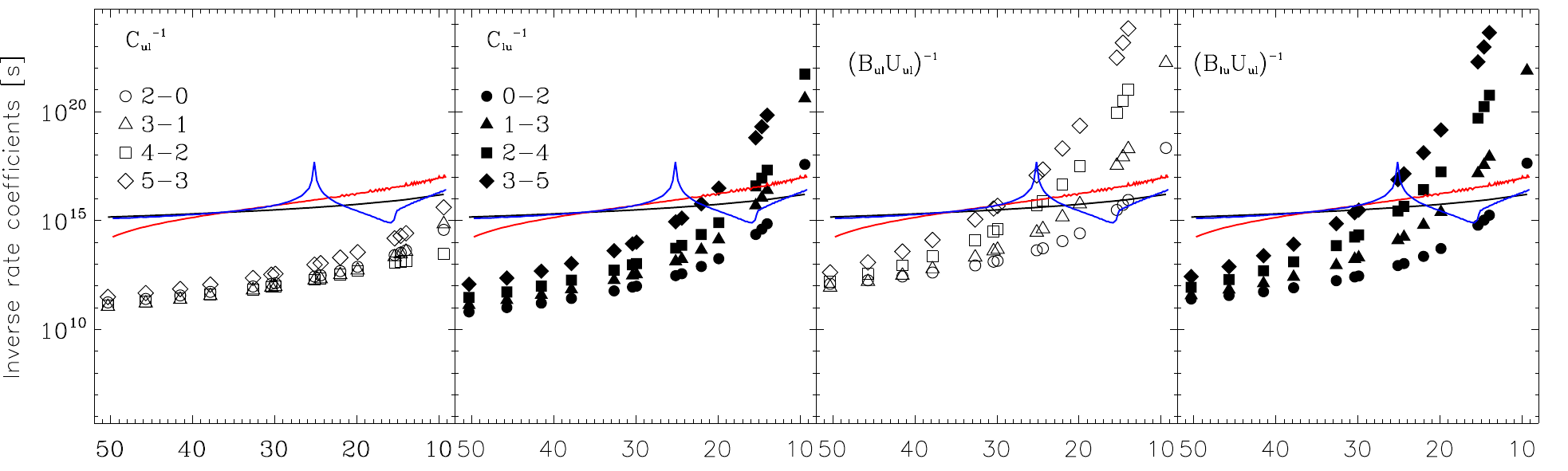}
\includegraphics[width=1.0\textwidth]{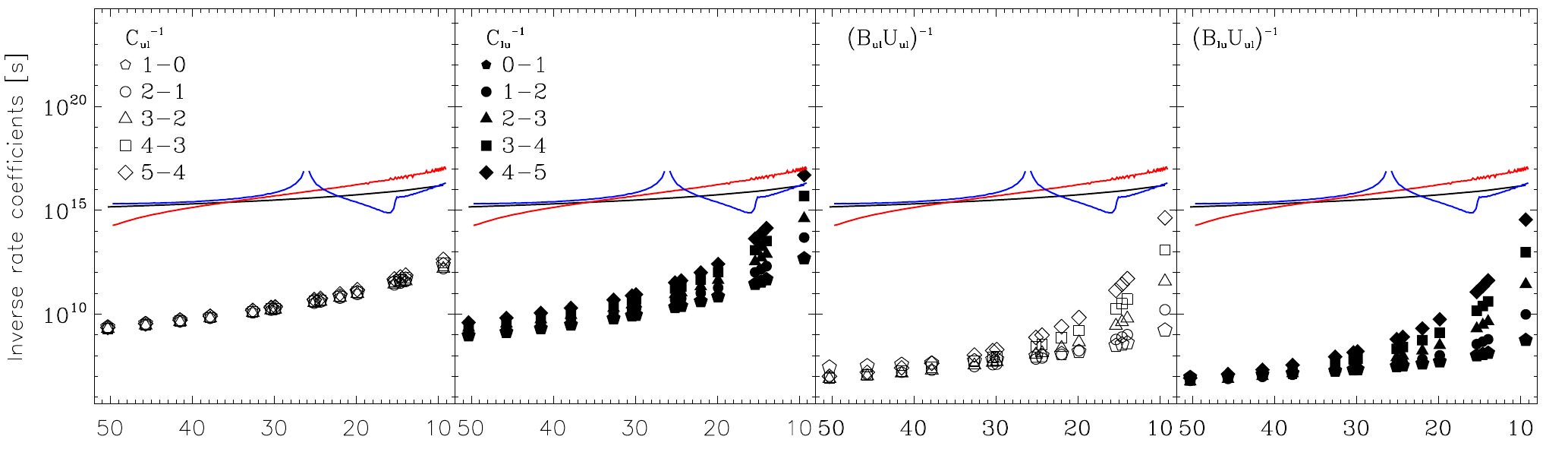}
\caption{The inverse rate coefficients $C_{ul}^{-1}$, $C_{lu}^{-1}$, $(B_{ul}U_{\nu_{ul}})^{-1}$ and $(B_{lu}U_{\nu_{ul}})^{-1}$ for molecules H$_2$ (top panel) and HD (bottom panel) in halos virialized at different $z$. The black solid line shows the age of the Universe corresponding to $z$, the red solid line shows the character time of number density change of molecules H$2$/HD in halo virialized at $z\approx50$ and  blue line at $z\approx15$.}
\label{inv_rates}
\end{figure*}

In Figure \ref{inv_rates} we show the inverse rates collision and radiative excitations, $C_{lu}^{-1}$ and $(B_{lu}U_{\nu_{ul}})^{-1}$, as well as de-excitations, $C_{ul}^{-1}$ and $(B_{ul}U_{\nu_{ul}})^{-1}$, in the halos virialized at different $z$ for comparison with actual age of the Universe $t_U$ and character time of number densities change of molecules, which is estimated as follows $\left(|d\ln{n_{X}}/dt|\right)^{-1}=\left(|(z+1)H(z)d\ln{n_{X}}/dz|\right)^{-1}$. Here and below $U_{\nu_{ul}}$ notes the energy density of CMB radiation at the frequency of transition between levels $u$ and $l$. One can see that in the dark ages halos for molecules H$_2$ the rates of collisional excitation/de-excitation by atomic hydrogen are comparable with the rates of radiative excitation/de-excitation by CMB quanta. At the high redshifts they are close to the value of rate of spontaneous transition $A_{ul}$ to the base level (Table \ref{Tab2}). For molecules HD in the conditions of dark ages halos the rates of collisional excitation/de-excitation are in few order lower than the radiative ones and the rates of spontaneous transitions  (Table \ref{Tab2}).

The values presented in Figure \ref{inv_rates} also show that for halos formed at $z>30$ ($T_K>400$ K, $n_H>10^6$ m$^{-3}$) the character times of collisional and radiation excitations/de-excitations of levels 2-5 of H$_2$ are lower than the age of the Universe and character time of number density change of molecular hydrogen. This means that the population of the rotational levels of molecular hydrogen can be estimated using condition of quasi-stationarity. Later, at $z\le30$, when adiabatic temperature of virialized halos is $<400$ K the character times of collisional and radiation excitations of levels 2-5 are comparable and higher than the age of the Universe, hence, the condition of quasi-stationarity for populations of the rotational levels there can not be used.  For molecules HD this condition can be used for dark ages halos practically always.

The presented here data about rates of spontaneous transitions allowed by quantum selection rules and rates of their collisional de-excitations say us about emission properties of halos in general. For any transition between levels $u$ and $l$, there is a critical number density of perturbers, $n^{cr}_H$, at which the radiating molecule suffers collisions at the rate $C_{ul}=A_{ul}$. It means that for $n_H\gg n^{cr}_H$ collisional de-excitations are dominated over spontaneous radiation transitions. In this case the thermal emission in the line vanishes. 

\begin{figure*}
\includegraphics[width=0.49\textwidth]{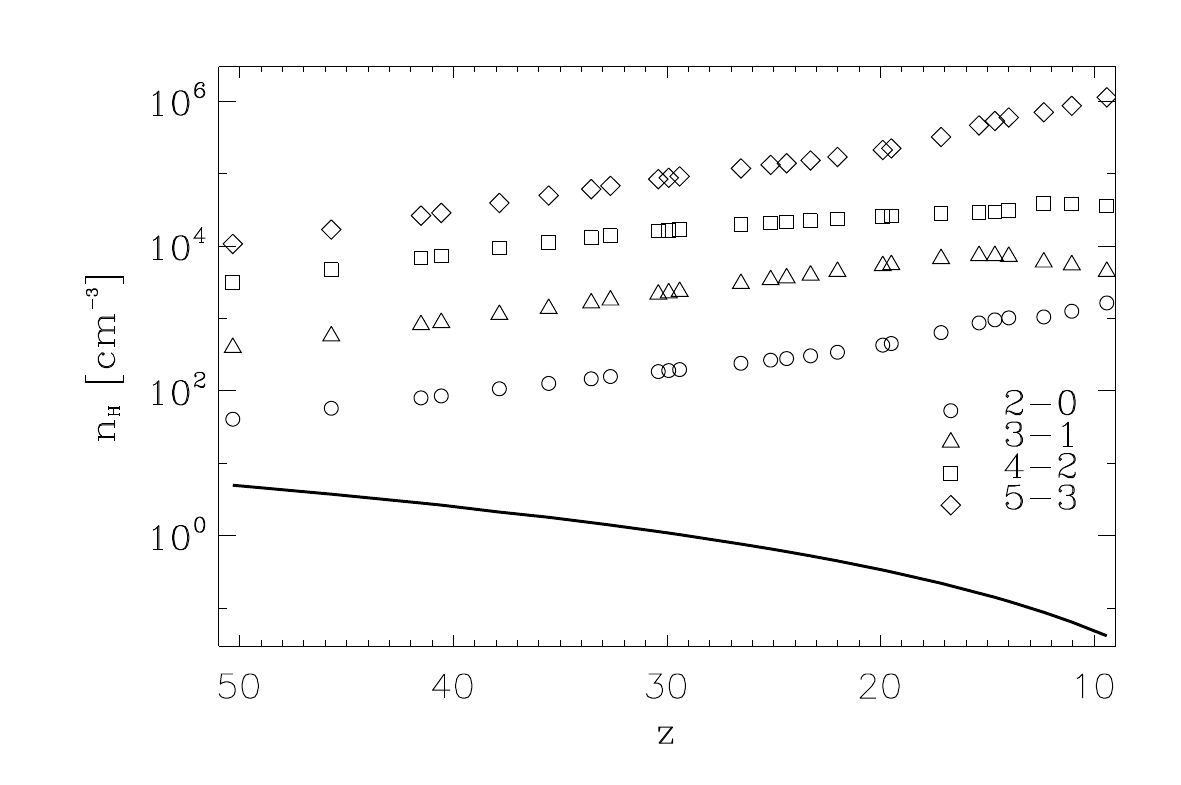}
\includegraphics[width=0.49\textwidth]{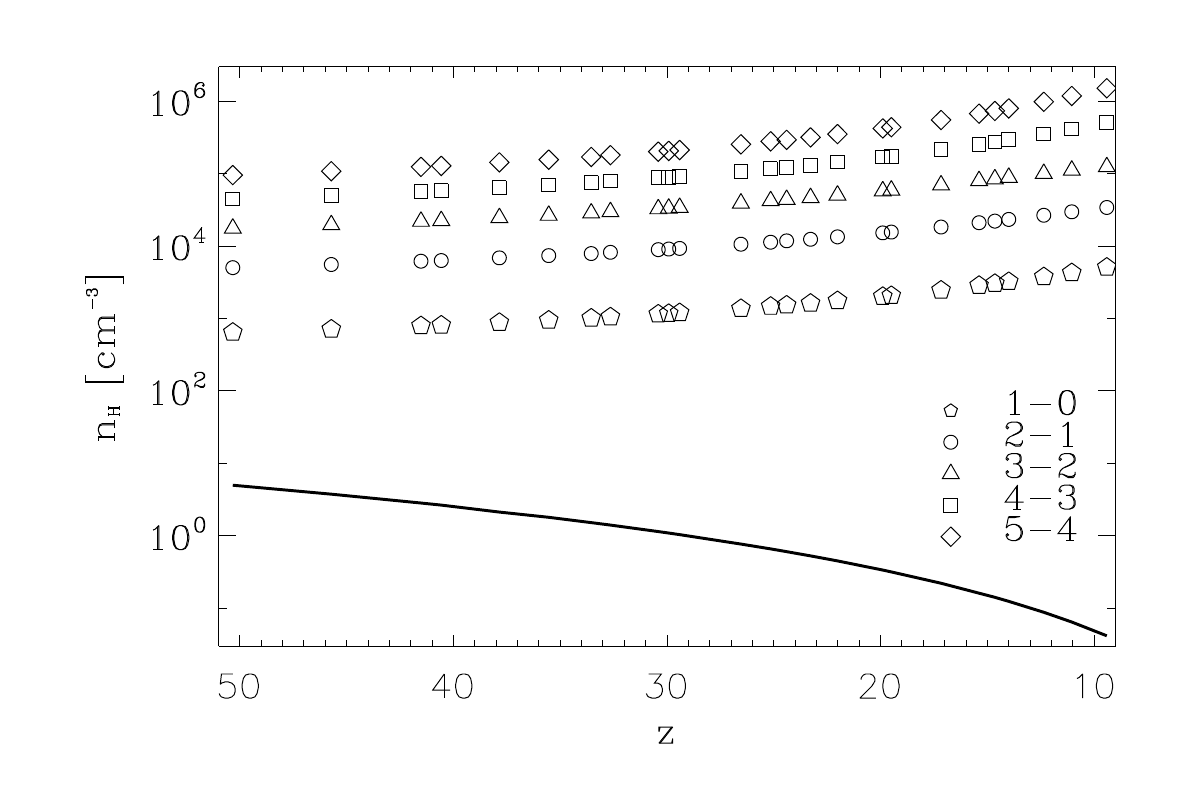}
\caption{The critical number density for lowest energy levels of molecular hydrogen (left panel) and hydrogen deuteride (right panel) in the dark ages halos virialized at different redshifts. The solid line shows the real values of number density of molecular hydrogen in the halos virialized at $z$. }
\label{cr_nH}
\end{figure*}

One can estimate the critical densities of atomic hydrogen for each level discussed here as 
$n_H^{cr}=A_{ul}/\kappa_{ul}$. The results of computations are presented in Figure \ref{cr_nH}. The critical densities for lowest rotational levels of molecular hydrogen H$_2$ and hydrogen deuteride HD are essentially larger than the density of the atomic hydrogen in the halos. It means that the thermal emission in these lines is expected. The problem is to extract it from the CMB background.


\section{Populations and excitation temperatures of rotational levels with $J=0-5$}

Let's consider the 6-levels system and all collision and radiative transition permitted by the quantum selection rules. 
Since the conditions of quasi-stationarity in the halos virialized at $z<30$ are not satisfied for all levels of molecule H$_2$ the kinetic equations for transitions between levels must be solved.
The selection rules for molecule H$_2$ (electric quadrupole transitions) permit $\Delta J=\pm2$ for radiative transitions, for molecule HD (electric dipole transitions) they permit $\Delta J=\pm1$, so the kinetic equations for populations of the rotational levels for these molecules are different.

In the general case before and after virialization the kinetic equations for populations of the rotational levels of H$_2$ and HD are as follows
\begin{widetext}
\begin{eqnarray}
 &&\hskip-0.5cm\rm{H_2:}\nonumber\\ 
 &&(1+z)H\frac{dX_j}{dz}=X_j\left(R_{j\,j+2}+C_{j\,j+4}\right)-X_{j+2}R_{j+2\,j}-X_{j+4}C_{j+4\,j}, \quad j=0,1\nonumber \\
 &&(1+z)H\frac{dX_j}{dz}=X_j\left(R_{j\,j-2}+R_{j\,j+2}\right)-X_{j-2}R_{j-2\,j}-X_{j+2}R_{j+2\,j}, \quad j=2,3\label{keH2}\\
 &&(1+z)H\frac{dX_j}{dz}=X_j\left(R_{j\,j-2}+C_{j\,j-4}\right)-X_{j-2}R_{j-2\,j}-X_{j-4}C_{j-4\,j}, \quad j=4,5\nonumber \\
 &&\hskip-0.5cm\rm{HD:}\nonumber\\ 
 &&(1+z)H\frac{dY_0}{dz}=Y_0\left(R_{01}+\sum_{i=2}^5C_{0i}\right)-Y_1R_{10}-\sum_{i=2}^5Y_iC_{i0}, \nonumber \\ 
 &&(1+z)H\frac{dY_j}{dz}=Y_j\left(R_{j\,j-1}+R_{j\,j+1}+\sum_{i\ne j,j\pm1}^5C_{ji}\right)-Y_{j-1}R_{j-1\,j}-Y_{j+1}R_{j+1\,j}-\sum_{i\ne j, j\pm1}^5 Y_iC_{ji}, \quad j=2,\,3,\,4\label{keHD}\\ 
 &&(1+z)H\frac{dY_5}{dz}=Y_5\left(R_{54}+\sum_{i=0}^3C_{5i}\right)-Y_4R_{45}-\sum_{i=0}^3Y_iC_{i5}, \nonumber 
\end{eqnarray}
\end{widetext}
where $X_j\equiv n_j/n_{H_2}$ and $Y_j\equiv n_j/n_{HD}$ are fractions of molecular hydrogen and hydrogen deuteride correspondingly in the state with rotational quantum number $J=j$, 
$R_{ul}=A_{ul}+B_{ul}U_{\nu_{ul}}+C_{ul}$, $R_{lu}=B_{lu}U_{\nu_{ul}}+C_{lu}$. One equation from system (\ref{keH2}) for H$_2$ can be substituted by simple equation
\begin{eqnarray}
\frac{n^{H_2}_1+n^{H_2}_3+n^{H_2}_5}{n^{H_2}_0+n^{H_2}_2+n^{H_2}_4}=\frac{n_{ortho}}{n_{para}},
\label{o-p}
\end{eqnarray}
and one equation from system (\ref{keHD}) for HD can be substituted by 
\begin{equation}
n^{HD}_0+n^{HD}_1+n^{HD}_2+n^{HD}_3+n^{HD}_4+n^{HD}_5=n_{HD}
\label{n_HD}
\end{equation}
after their differentiation with respect to redshift.

We integrate the equations (\ref{keH2})-(\ref{keHD}) by the code {\it\large ddriv1}\footnote{http://www.netlib.org/slatec/src/ddriv1.f} together with equations of evolution of temperature and density of all components as well as with kinetic equations of formation/destruction of molecules in the halo \citep{Novosyadlyj2018}. 
We set the initial conditions for eqs. (\ref{keH2})-(\ref{keHD}) assuming that at $z\ge200$ the quasi-stationary condition is satisfied and we solve the system of algebraic equations for H$_2$ and HD separately. In such way we obtain the evolution of populations of the rotational levels, opacities and the line intensities at all stages formation and existence of dark ages halos.

Dark ages halos after virialization have unchanged density, temperature and chemical composition, hence, the conditions of stationarity of populations  of energy levels are satisfied for them. In this case the left
parts of eqs. (\ref{keH2})-(\ref{keHD}) are zero, 
$$\frac{dX_j}{dz}=0, \quad \frac{dY_j}{dz}=0,$$ 
and we have two systems of algebraic equations (\ref{keH2}) and (\ref{keHD}) which can be solved by standard methods if their last equations, for example, will be substituted by equations (\ref{o-p})-(\ref{n_HD}) correspondingly. We use for that the subroutine dgesv.f from lapack library. 
\begin{table}[ht!]
\begin{center}
\caption{The excitation temperatures for lowest rotational levels of H$_2$-molecules in the halos formed at the redshifts $z=30-10$.}
\begin{tabular} {lcccccc}
\hline
\hline
   \noalign{\smallskip}
$z$&$T_K$ [K]&$T_R$ [K]&\multicolumn{4}{c}{$T_{ex}$ [K]}  \\
 \noalign{\smallskip} 
& & &0-2&1-3&2-4&3-5\\
 \noalign{\smallskip} 
\hline
   \noalign{\smallskip} 
30.41&402.1&85.6&93.7&93.3&113.4&145.7 \\
    \noalign{\smallskip}
25.15&298.9&71.3& 76.1& 77.5& 102.9& 130.3 \\ 
    \noalign{\smallskip}
19.90&206.3&57.0&59.7& 62.7& 89.5& 108.6 \\ 
    \noalign{\smallskip}
14.65&124.3&42.7& 43.8& 48.1& 69.9& 79.4 \\  
 \noalign{\smallskip}
9.41& 59.8&28.4& 28.8& 32.9& 43.7& 44.6 \\ 
 \noalign{\smallskip} 
  \hline
\end{tabular}
\label{Tab3}
\end{center}
\end{table}
\begin{table}[ht!]
\begin{center}
\caption{The excitation temperatures for lowest rotational levels of molecules HD in the halos formed at the redshifts $z=30-10$.}
\begin{tabular} {lcccccc}
\hline
\hline
   \noalign{\smallskip}
$z$&$T_K$ [K]&$T_R$ [K]&\multicolumn{4}{c}{$T_{ex}$ [K]} \\
 \noalign{\smallskip} 
& & &0-1&1-2&2-3&3-4 \\
 \noalign{\smallskip} 
\hline
   \noalign{\smallskip} 
30.41&402.1&85.61&85.71&85.64&85.63&85.63 \\
    \noalign{\smallskip}
25.15&298.9&71.27& 71.32& 71.29& 71.28& 71.28 \\ 
    \noalign{\smallskip}
19.90&206.3&56.96&56.98& 56.97& 56.97& 56.97 \\ 
    \noalign{\smallskip}
14.65&124.3&42.65& 42.66& 42.66& 42.66& 42.66 \\ 
 \noalign{\smallskip}
 9.41& 59.8&28.37& 28.37& 28.37& 28.37& 28.37 \\ 
 \noalign{\smallskip} 
  \hline
\end{tabular}
\label{Tab3a}
\end{center}
\end{table}

When the populations of the rotational levels are computed, we can calculate the excitation temperatures for $l-u$ levels: 
$$T_{ex}=\frac{h\nu_{ul}}{k_B}\left[\ln{\frac{g_un_l}{g_ln_u}}\right]^{-1}.$$
They are presented for for $l-u$ levels of molecules H$_2$ and HD in Tables \ref{Tab3} and \ref{Tab3a} accordingly for the dark ages halos with mass $M_h=5.3\cdot10^9$ M$_{\odot}$ and different initial amplitude of matter density perturbations, which are virialized at $z=$30.41, 25.15, 19.90, 14.65 and 9.41. The kinetic temperature of gas in halos and CMB temperature are presented for comparison. One can see that values of $T_{ex}-T_R>0$, so, the halos are warmer spots in the cosmic microwave background. Note, that these values are essentially larger for molecules H$_2$ than for HD. It is because the hydrogen deuteride molecule are asymmetrical, has nonzero electric dipole moment and more effectively interacts with radiation. The data in Table \ref{Tab2} and in Figure \ref{inv_rates} support this explanation.

The excitation temperatures of the lowest rotational levels of molecular hydrogen are sensitive to the ratio of $n_{ortho}/n_{para}$ and values of rate coefficients of their collisional excitation/de-excitation by atomic hydrogen. We re-compute the transitions ortho$\rightleftarrows$para and populations of the rotational levels using  the rate coefficients revised by \cite{Lique2015}. It resulted in some differences with the data of \cite{Flower2000}: in the temperature range 1000-10 K, the ratio $n_{ortho}/n_{para}$ changes in the range 1-1.3, instead of 2.8-0.3. The excitation temperatures, however, do not change so radically: differences do not exceed 4\% for the first excited levels of para- and ortho-H$_2$ with $J=2$ and $J=3$ accordingly. But the populations of these levels changed more significantly: in the temperature range of $\sim$400-60 K the population of the first excited level of para-H$_2$ increases by 1.7-2.2 times, while the population of the first excited level of ortho-H$_2$ decreases by 1.6-3.2 times.

\section{Opacity and brightness temperature of dark ages halos in rotational lines of molecules H$_2$ and HD}

The opacity at the frequency of $u-l$ transition is calculated in the rest frame of the halo as
$$\tau_{ul}=\int_0^{r_h} \alpha_{\nu_{ul}}(r)dr,$$
where $\alpha_{\nu_{ul}}$ is the absorption coefficient per unit of length in frequency $\nu_{ul}$, and $r_h$ is radius of halo. 

The absorption coefficient per unit of length in frequency $\nu_{ul}$ is as follows (formulas (2.69) and (2.154) in \citet{Lang1974}):
\begin{equation}
\alpha_{\nu_{ul}}=\frac{c^2}{8\pi\nu_{ul}^2}\frac{n_l}{\Delta\nu_L}\frac{g_u}{g_l}\left[1-\exp{\left(-\frac{h\nu_{ul}}{k_B T_{ex}}\right)}\right]A_{ul},
\label{alpha_ul}
\end{equation}
where $n_l$ is the number density of molecules in the lower level $l$, $k_B$ is Boltzmann constant, $A_{ul}$ is the spontaneous transition probability
(Einstein coefficient), $T_{ex}$ is excitation temperature for $u-l$ transition, $\Delta\nu_L$ is the width of line at a half-maximum. The last value in the case of the virialized halo can be presented like in the case of turnaround halo \citep{Maoli1996}): 
\begin{equation}
\frac{\Delta\nu_L}{\nu}=\frac{2}{c}\sqrt{\frac{2\ln{2}\,k_B T_K}{m}}=7.16\cdot10^{-7}\sqrt{\frac{T_K}{m_A}},
\end{equation}
where $T_K$ is kinetic temperature (adiabatic or virial) of gas in halo, $m$ and $m_A$ are atomic mass and atomic number of a molecule correspondingly. 
In the visible center of the spherical homogeneous top-hat halo the optical depth in rotational line $\nu_{ul}$ is as follows
\begin{equation}
\tau_{ul}=1.55\cdot10^{50}n_l\frac{g_u}{g_l}\frac{A_{ul}}{\nu_{ul}^3}\sqrt{\frac{m_A}{T_k}}\left[1-\exp{\left(-\frac{h\nu_{ul}}{k_B T_{ex}}\right)}\right]r_h,
\label{tau}
\end{equation}
where $r_h$ is its radius in units of Mpc (the rest values must be in CGS system).

\begin{table}[ht!]
\begin{center}
\caption{The opacity of halos with $M_h=5.3\cdot10^9$ M$_\odot$ virialized at different redshifts $z_v$ in the lowest rotational levels of molecular hydrogen H$2$.}
\begin{tabular} {lcccc}
\hline
\hline
   \noalign{\smallskip}
$z_v$ &\multicolumn{4}{c}{$\tau_{ul}$}\\
 \noalign{\smallskip} 
&2-0&3-1&4-2&5-3\\
 \noalign{\smallskip} 
\hline
   \noalign{\smallskip} 
30.41 &8.67$\cdot10^{-9}$&   4.24$\cdot10^{-8}$&   5.23$\cdot10^{-10}$&   2.94$\cdot10^{-11}$\\
25.15 &4.99$\cdot10^{-9}$&   2.48$\cdot10^{-8}$&   8.55$\cdot10^{-11}$&   2.74$\cdot10^{-12}$\\
19.90 &2.67$\cdot10^{-9}$&   1.35$\cdot10^{-8}$&   7.26$\cdot10^{-12}$&   1.13$\cdot10^{-13}$\\
14.65 &1.43$\cdot10^{-9}$&   6.64$\cdot10^{-9}$&   1.76$\cdot10^{-13}$&   9.45$\cdot10^{-16}$\\
 9.41 &9.40$\cdot10^{-10}$&  2.60$\cdot10^{-9}$&   2.72$\cdot10^{-16}$&   1.09$\cdot10^{-19}$\\ 
 \noalign{\smallskip} 
  \hline
\end{tabular}
\label{Tab_tau_H2}
\end{center}
\end{table}

\begin{table}[ht!]
\begin{center}
\caption{The opacity of halos with $M_h=5.3\cdot10^9$ M$_\odot$ virialized at different redshifts $z_v$ in the lowest rotational levels of hydrogen deuteride molecules HD.}
\begin{tabular} {ccccc}
\hline
\hline
   \noalign{\smallskip}
$z_v$ &\multicolumn{4}{c}{$\tau_{ul}$}\\
 \noalign{\smallskip} 
&1-0&2-1&3-2&4-3 \\
 \noalign{\smallskip} 
\hline
   \noalign{\smallskip} 
30.41 &2.25$\cdot10^{-7}$&   1.23$\cdot10^{-7}$&   9.69$\cdot10^{-9}$&   1.51$\cdot10^{-10}$ \\
25.15 &3.17$\cdot10^{-7}$&   1.22$\cdot10^{-7}$&   5.18$\cdot10^{-9}$&   3.28$\cdot10^{-11}$ \\
19.90 &4.28$\cdot10^{-7}$&   9.93$\cdot10^{-8}$&   1.69$\cdot10^{-9}$&   2.78$\cdot10^{-12}$ \\
14.65 &3.12$\cdot10^{-7}$&   3.23$\cdot10^{-8}$&   1.21$\cdot10^{-10}$&  2.09$\cdot10^{-14}$ \\
 9.41 &1.25$\cdot10^{-7}$&   2.74$\cdot10^{-9}$&   4.98$\cdot10^{-13}$&  9.58$\cdot10^{-19}$ \\ 
 \noalign{\smallskip} 
  \hline
\end{tabular}
\label{Tab_tau_HD}
\end{center}
\end{table}

The results of computations of $\tau_{ul}$ for the lowest rotational levels of molecules para-/ortho-H$_2$ and HD are presented in Tables \ref{Tab_tau_H2} and \ref{Tab_tau_HD} accordingly.
The opacities in the frequency $\nu_{31}$ of ortho-H$_2$ molecule is larger than in the frequency $\nu_{20}$ of para-H$_2$ molecule because $n_{ortho}/n_{para}\approx3$ \citep{Flower2000}. They decrease with decreasing of $z_v$ because the number density of molecules decreases (Table \ref{Tab1})  as well as the populations of non-base levels of both molecules via decreasing of excitation temperatures (Table \ref{Tab3} and \ref{Tab3a}). The opacity related with the base level of HD molecule is not monotonic function of $z_v$ (2nd column of Table \ref{Tab_tau_HD}) since its population ($n_{J=0}/n_{HD}$) fast increases with decreasing of excitation temperature.

We note also, that opacities of the dark ages halos in the rotational lines of molecule HD are higher than in the lines of H$_2$ in spite of the $n_{HD}/n_{H_2}\sim10^{-4}$. It has simple explanation: the factor $A_{ul}/\nu^2_{ul}$ in (\ref{alpha_ul}) is essentially larger for HD than for H$_2$.

\begin{figure*}[!ht]
\includegraphics[width=0.49\textwidth]{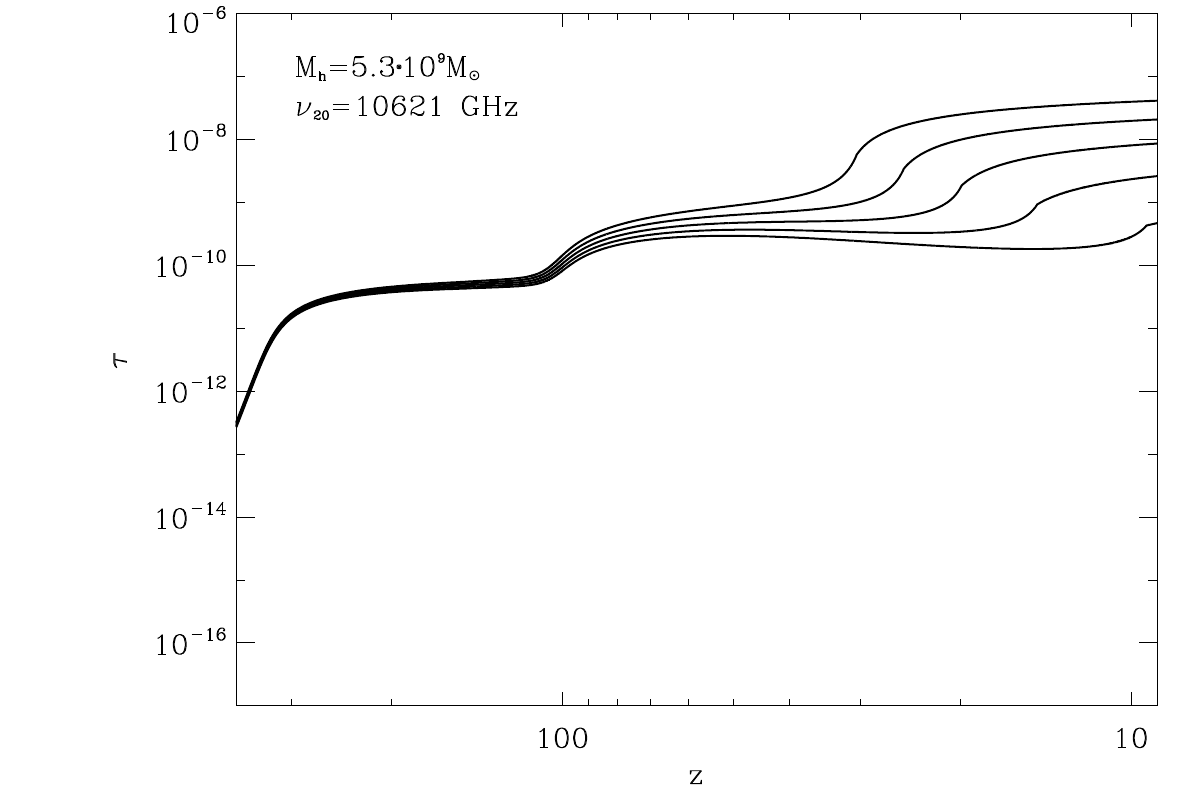}
\includegraphics[width=0.49\textwidth]{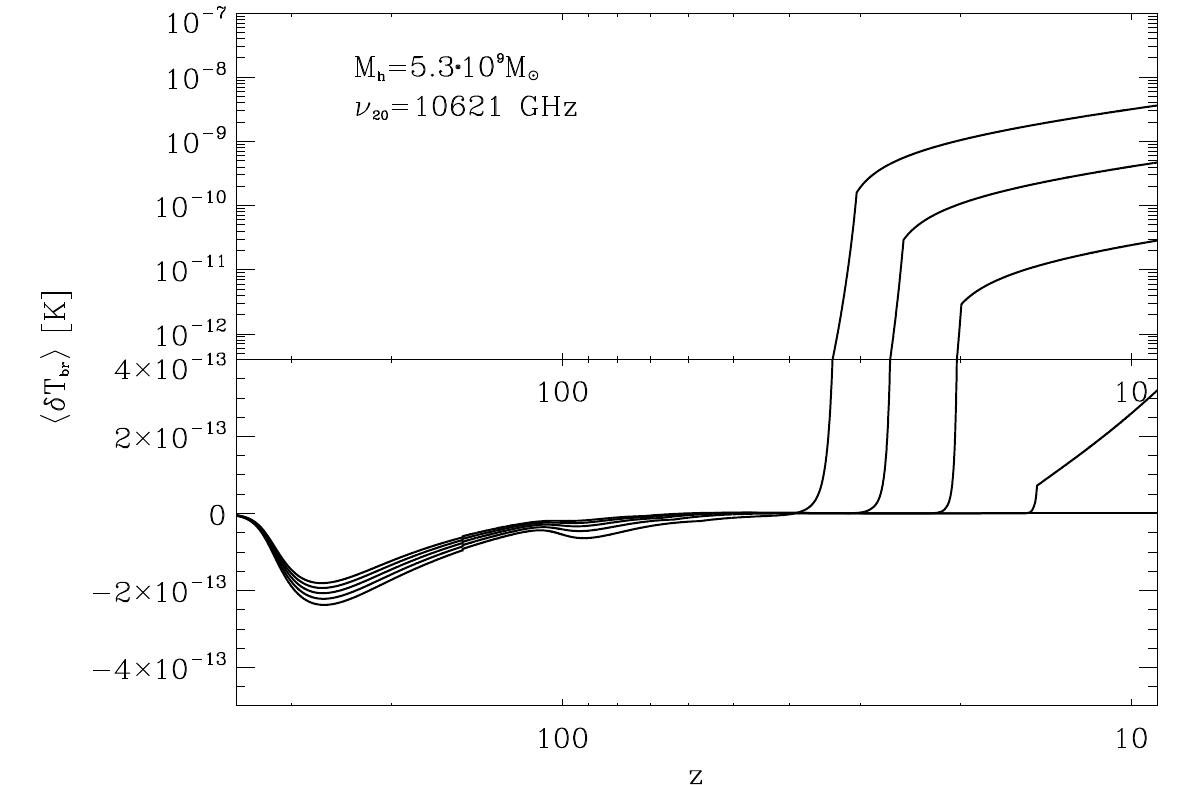}
\includegraphics[width=0.49\textwidth]{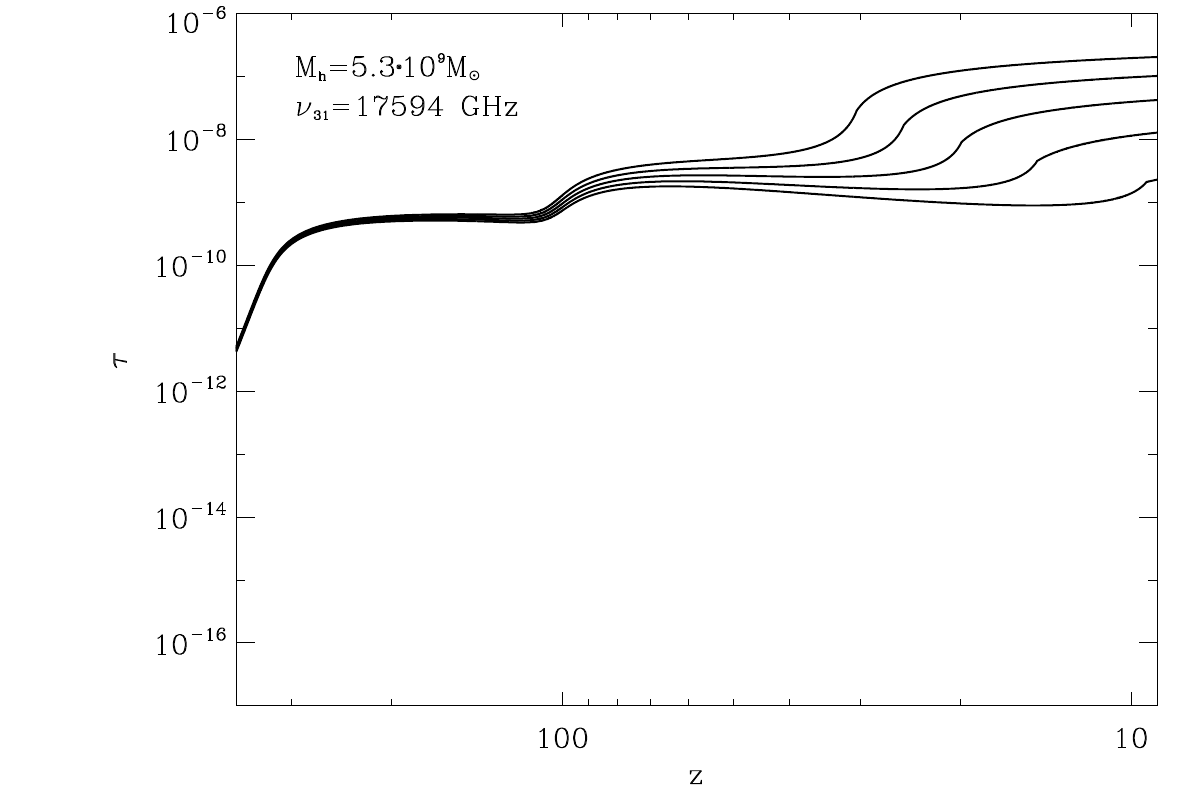}
\includegraphics[width=0.49\textwidth]{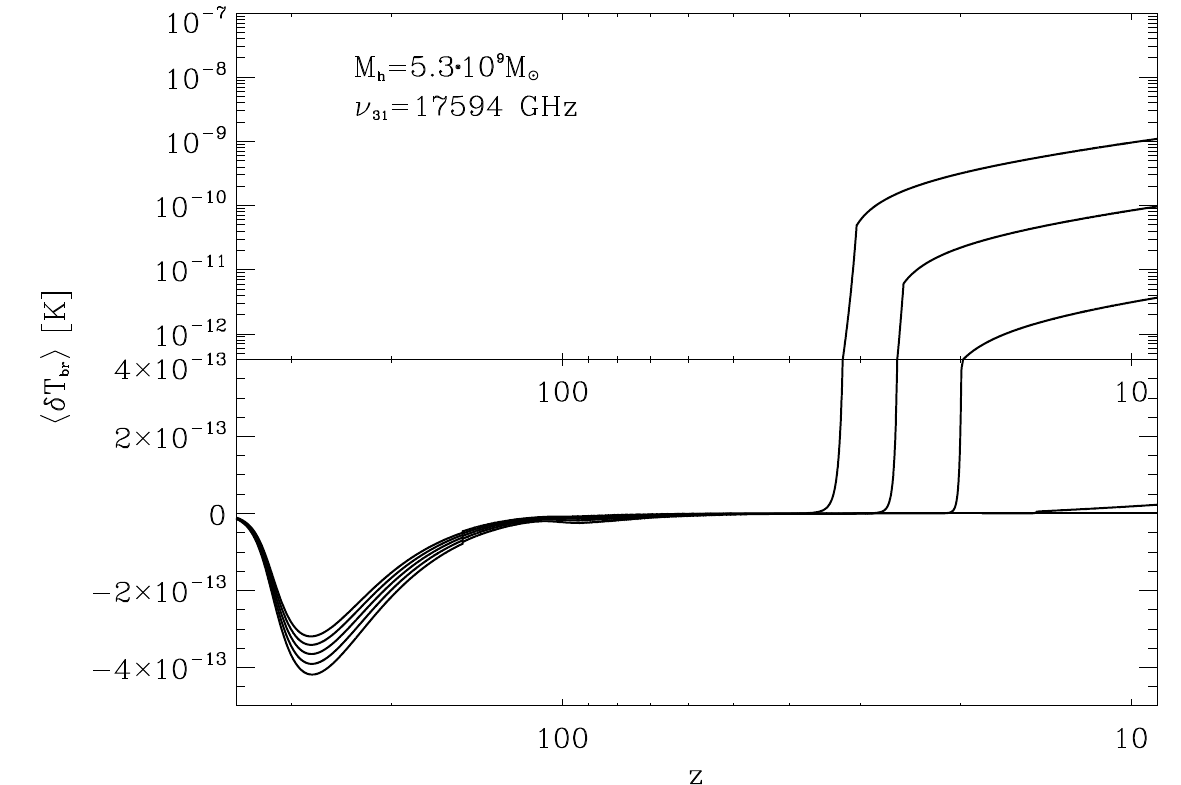}
\caption{Evolution of opacity (left column) and brightness temperature (right column) in the lines of transitions $J=2\rightarrow0$ (top row) and $J=3\rightarrow1$ (bottom row) of molecule H$_2$ for halos with mass $M_h=5.3\cdot10^9$ M$_\odot$. Each line corresponds to the halo with different initial amplitude of curvature perturbation: $C_k=3\cdot10^{-4},\,2.5\cdot10^{-4},\,2\cdot10^{-4},\,1.5\cdot10^{-4},\,1\cdot10^{-4}$ (from top to bottom in the right hand side of each panel. }
\label{Tbe_H2}
\end{figure*}

\begin{figure*}[!ht]
\includegraphics[width=0.49\textwidth]{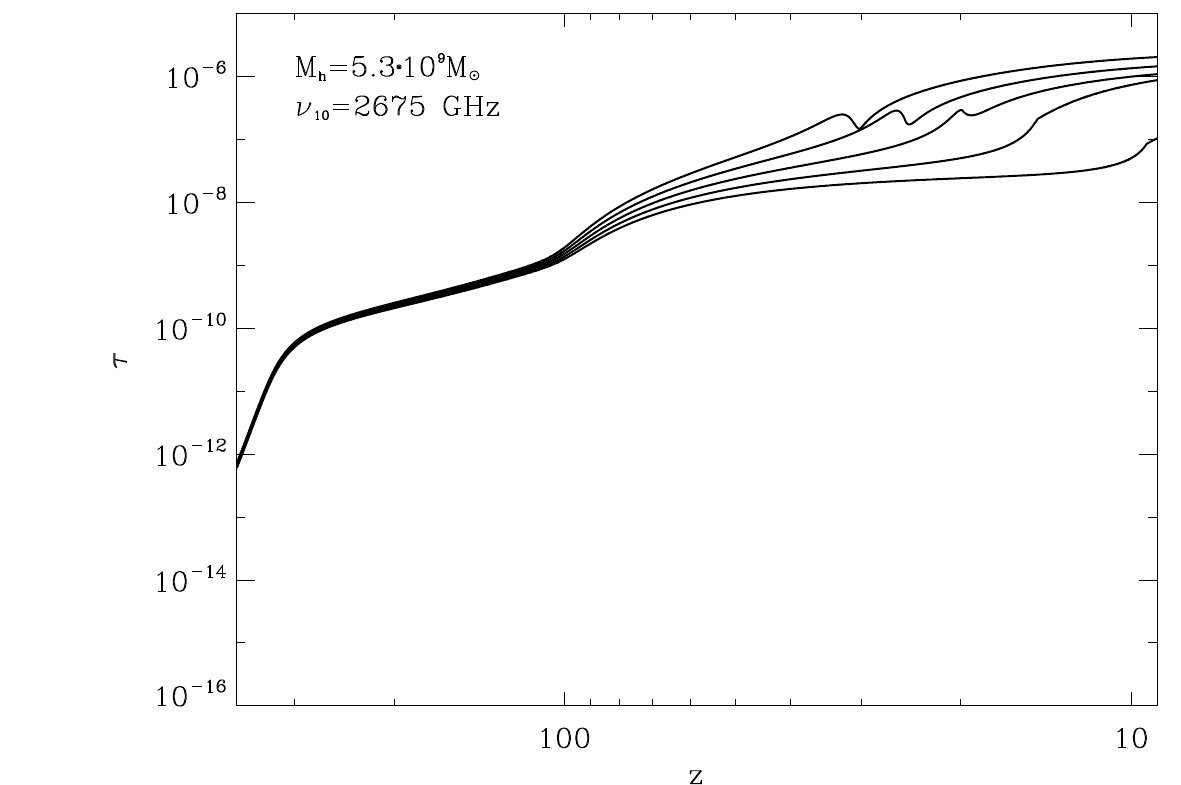}
\includegraphics[width=0.49\textwidth]{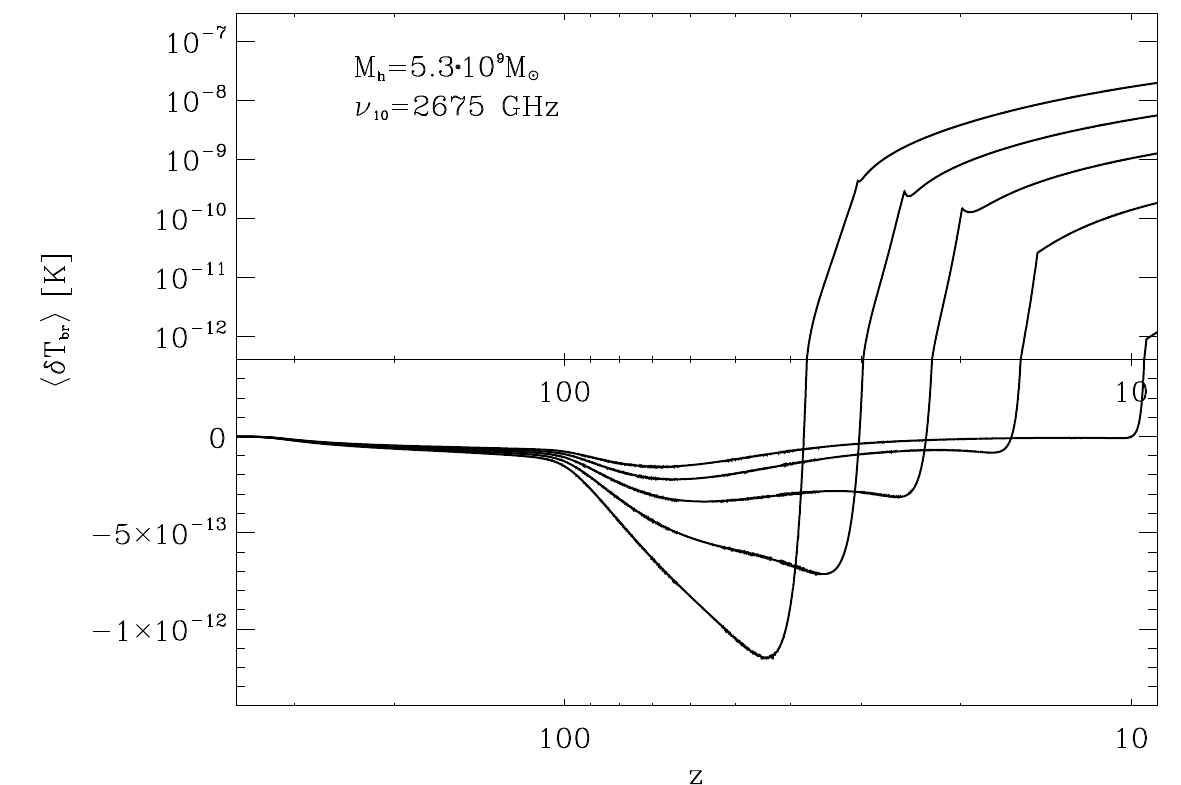}
\caption{Evolution of of opacity (left column) and brightness temperature (bright column) in the lines of transitions $J=1\rightarrow0$  of molecule HD for halos with mass   $M_h=5.3\cdot10^9$ M$_\odot$. Each line corresponds to the halo with different initial amplitude of curvature perturbation as in the previous figure. }
\label{Tbe_HD}
\end{figure*}

We also have analyzed the evolution of opacities of single halos during its formation. The results for the brightest lines of H$_2$ and HD molecules for halos with mass $M_h=5.3\cdot10^9$ M$_\odot$ and different initial amplitudes of density perturbation are presented in the left panels of Figures \ref{Tbe_H2} and \ref{Tbe_HD} accordingly. The same results for other lines and other halos are presented in the animate figures 8 and 9 in Appendix of the on-line version. One can see that halos formed earlier have larger opacity at lower redshifts in comparison with halos formed at the same redshifts, since they are denser. 


The differential brightness temperature for thermal emission is obtained from the radiative transfer equation and is as follows 
\begin{equation}
\delta T^b_{ul}=\frac{h\nu_{obs}}{k_B}\left[\frac{1}{e^{\frac{h\nu_{ul}}{k_BT_{ex}}}-1} -\frac{1}{e^{\frac{h\nu_{ul}}{k_BT_R}}-1}\right]\left(1-e^{-\tau_{ul}}\right),\nonumber
\end{equation}
where $T_b$ is the Rayleigh-Jeans brightness temperature, $T_{ex}$ is the excitation temperature of a transition $u-l$ and $T_R$ is temperature of the background radiation. The dark ages halos are optically thin in the molecular lines, so, $(1-e^{-\tau_{ul}})\approx \tau_{ul}$, and the expression for the brightness temperature of thermal emission of dark ages halos becomes as follows
\begin{equation}
\delta T^b_{ul}=\frac{7.44\cdot10^{39}n_uA_{ul}}{(1+z)\nu_{ul}^2}\sqrt{\frac{m_A}{T_K}}\left[\frac{e^{\frac{h\nu_{ul}}{k_BT_R}}-e^{\frac{h\nu_{ul}}{k_BT_{ex}}}}{e^{\frac{h\nu_{ul}}{k_BT_R}}-1}\right]\frac{r_h}{1\,\rm{Mpc}},
\label{dTb2}
\end{equation}
where $n_u$ is in units cm$^{-3}$. Since the observable differential brightness temperature is not homogeneous in the surface of halo we can average it like 
$$\langle\delta T_{br}\rangle=2\int_0^1\delta T_{br}(x)xdx,$$
where $x=\theta/\theta_h$. In our case the integral has exact analytical presentation which for small values of $\tau_{ul}$ gives factor 2/3:
\begin{eqnarray}
\langle\delta T^b_{ul}\rangle=\frac{2}{3}\delta T^b_{ul}.
\label{dTb3}
\end{eqnarray}
The expression (\ref{dTb2}) gives us the possibility to understand when thermal emission signal from dark ages halos can be sighted in the cosmic microwave background. For that the additional perturbers are necessary in order $T_{ex}>T_R$. If $T_{ex}\rightarrow T_R$ then $\delta T^b_{ul}\rightarrow0$. The halos should be dense and extended to have large opacities. The emitting molecules must be abundant and have a large value of ratio of Einstein coefficients to square frequency for excited levels. The maximal value of differential brightness temperature is expected in the case when $T_{ex}\gg h\nu_{ul}/k_B$ and value in brackets in (\ref{dTb2}) is close to 1. 

We use the data from Tables \ref{Tab1}, \ref{Tab1A} and \ref{Tab2} for computation the differential brightness temperatures in the lowest rotational frequencies of molecules H$_2$ and HD for halos formed at $z=10-50$. The results for halos with $M_h=5.3\cdot10^9$ M$_{\odot}$ are presented in Tables \ref{Tab_dTb_H2} and \ref{Tab_dTb_HD} for molecules H$_2$ and HD accordingly.
\begin{table}[ht!]
\begin{center}
\caption{Thermal emission: the differential brightness temperatures of halos with $M_h=5.3\cdot10^9$ M$_\odot$ virialized at different redshifts $z_v$ in the lowest rotational levels of molecular hydrogen H$2$.}
\begin{tabular} {lcccc}
\hline
\hline
   \noalign{\smallskip}
$z_v$ &\multicolumn{4}{c}{$\langle\delta T^b_{ul}\rangle$ [K]}\\
 \noalign{\smallskip} 
&2-0&3-1&4-2&5-3\\
 \noalign{\smallskip} 
\hline
   \noalign{\smallskip} 
30.41 &1.64$\cdot10^{-10}$&  4.92$\cdot10^{-11}$&  4.08$\cdot10^{-13}$&  3.40$\cdot10^{-14}$ \\
25.15 &2.92$\cdot10^{-11}$&  6.13$\cdot10^{-12}$&  2.87$\cdot10^{-14}$&  1.12$\cdot10^{-15}$ \\
19.90 &2.87$\cdot10^{-12}$&  3.82$\cdot10^{-13}$&  5.61$\cdot10^{-16}$&  6.03$\cdot10^{-18}$ \\
14.65 &7.57$\cdot10^{-14}$&  5.13$\cdot10^{-15}$&  4.63$\cdot10^{-19}$&  4.33$\cdot10^{-22}$ \\
 9.41 &1.55$\cdot10^{-16}$&  9.80$\cdot10^{-19}$&  4.53$\cdot10^{-26}$&  3.39$\cdot10^{-32}$ \\ 
 \noalign{\smallskip} 
  \hline
\end{tabular}
\label{Tab_dTb_H2}
\end{center}
\end{table}
\begin{table}[ht!]
\begin{center}
\caption{Thermal emission: the differential brightness temperatures of halos with $M_h=5.3\cdot10^9$ M$_\odot$ virialized at different redshifts $z_v$ in the lowest rotational levels of hydrogen deuteride molecules HD.}
\begin{tabular} {ccccc}
\hline
\hline
   \noalign{\smallskip}
$z_v$ &\multicolumn{4}{c}{$\langle\delta T^b_{ul}\rangle$ [K]}\\
 \noalign{\smallskip} 
&1-0&2-1&3-2&4-3 \\
 \noalign{\smallskip} 
\hline
   \noalign{\smallskip}  
30.41 &4.28$\cdot10^{-10}$&   5.32$\cdot10^{-11}$&   2.02$\cdot10^{-12}$&   3.68$\cdot10^{-14}$ \\
25.15 &2.84$\cdot10^{-10}$&   2.29$\cdot10^{-11}$&   4.66$\cdot10^{-13}$&   4.50$\cdot10^{-15}$ \\
19.90 &1.45$\cdot10^{-10}$&   6.08$\cdot10^{-12}$&   4.94$\cdot10^{-14}$&   2.00$\cdot10^{-16}$ \\
14.65 &2.59$\cdot10^{-11}$&   3.44$\cdot10^{-13}$&   5.98$\cdot10^{-16}$&   6.11$\cdot10^{-19}$ \\
 9.41 &9.01$\cdot10^{-13}$&   9.67$\cdot10^{-16}$&   6.33$\cdot10^{-20}$&   3.60$\cdot10^{-24}$ \\ 
 \noalign{\smallskip} 
  \hline
\end{tabular}
\label{Tab_dTb_HD}
\end{center}
\end{table}
First of all we note that thermal emission of dark ages halos in the low rotational level of molecules H$_2$ and HD exist but the value of differential brightness temperatures are too low to be detected by current instrumentation. They are larger for halos virialized earlier and vanishes for halos virialized late. 

We also have analyzed the evolution of differential brightness temperature in the rotational lines of H$_2$ and HD of single halos during its formation. The results are shown for the lowest transitions 
($J=2\rightarrow0/J=3\rightarrow1$ for para-/ortho-H$_2$ and $J=1\rightarrow0$ HD molecules) in the right panels of Figures \ref{Tbe_H2} and \ref{Tbe_HD} for halos with mass $M_h=5.3\cdot10^9$ M$_\odot$ and for other transitions and halos of other mass are presented in the animate figures 8 and 9 in Appendix of the on-line version.  
These figures are combination of logarithmic and normal scales since before the turn around of halos and even some time of the beginning of the collapse the differential brightness temperature are negative since $T_{ex}<T_R$ because the adiabatic temperature of gas is lower than CMB temperature, $T_K<T_R$. In the case of molecule HD we see the deep lowering of the brightness temperature for transition $J=1\rightarrow0$, slight decrease  for transition $J=2\rightarrow1$ and its absence for transitions between upper levels before appearance of emission. Their amplitudes, however, are too small to discuss the possibility of their observations.

Another measure of the halo signal extracted from the cosmic background radiation is the differential energy flux per unit frequency \citep{Iliev2002}
\begin{eqnarray}
\delta F_{ul}&=&\frac{2\pi}{c}\left(\frac{\nu_{ul}}{1+z}\right)^2k_B\langle\delta T^b_{ul}\rangle\theta_h^2\\
\hskip1cm &=& \frac{2.27}{(1+z)^2}\left(\frac{\nu_{ul}}{1\,\rm{GHz}}\right)^2\left(\frac{\langle\delta T^b_{ul}\rangle}{1\,\rm{K}}\right)\left(\frac{\theta_h}{1''}\right)^2\,\,\rm{\mu Jy}\nonumber
\end{eqnarray}  
For the values of differential brightness temperatures presented in Tables \ref{Tab_dTb_H2} and \ref{Tab_dTb_HD}
the values of differential energy flux $\delta F_{ul}\le10^{-5}\,\rm{\mu Jy}$. In Tables \ref{Tab10} and \ref{Tab11} we present them for the rest halos and two lowest transitions of molecules H$_2$ and HD accordingly. The maximal values of differential energy fluxes we obtain for halo of maximal mass ($M_h=5.3\cdot10^9$ M$_\odot$) which virialized earliest ($z_v=30.41$). For example, in the frequency 338 GHz (transition $J=2\rightarrow0$ of para-H$_2$ molecule) it is  $\sim4\cdot10^{-5}\,\rm{\mu Jy}$. The maximal flux in the frequency 85 GHz ($J=1\rightarrow0$) of the molecule HD is estimated as $\sim7\cdot10^{-6}\,\rm{\mu Jy}$. Both values are essentially lower of the sensitivity of current radio telescopes. 

\begin{figure}[!ht]
\includegraphics[width=0.45\textwidth]{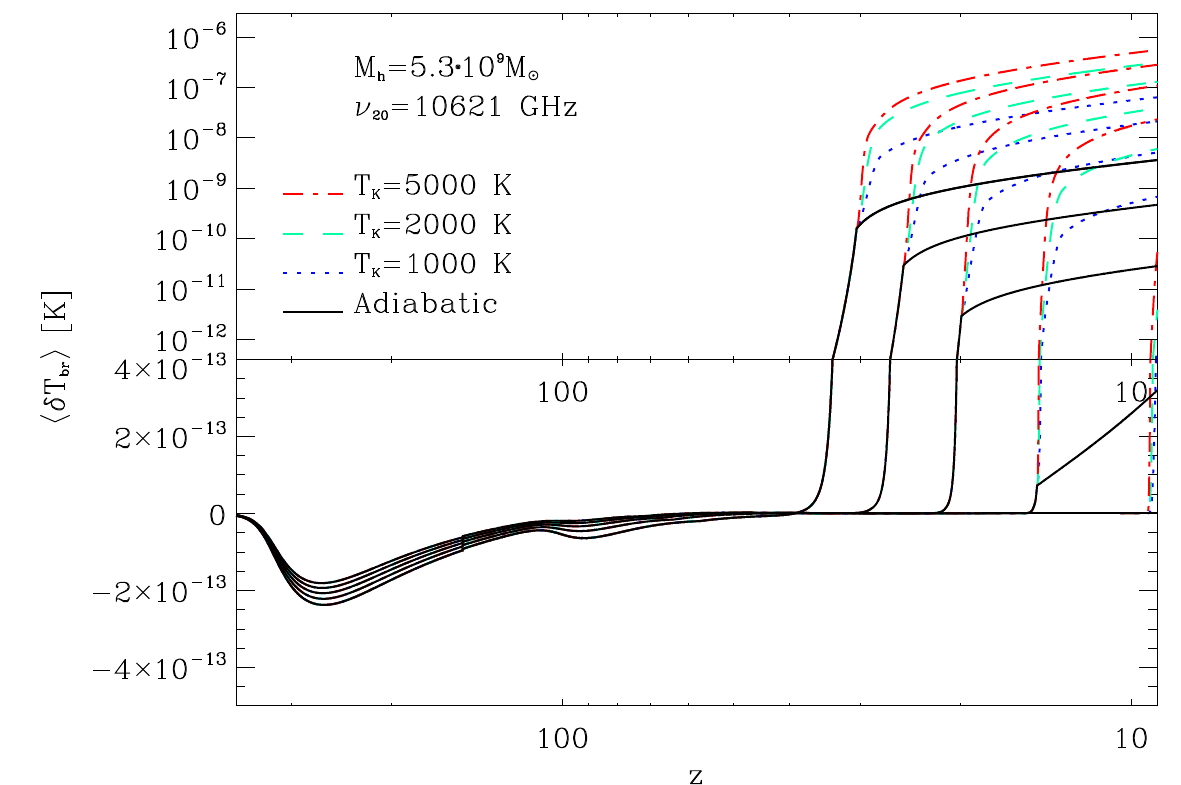}
\includegraphics[width=0.45\textwidth]{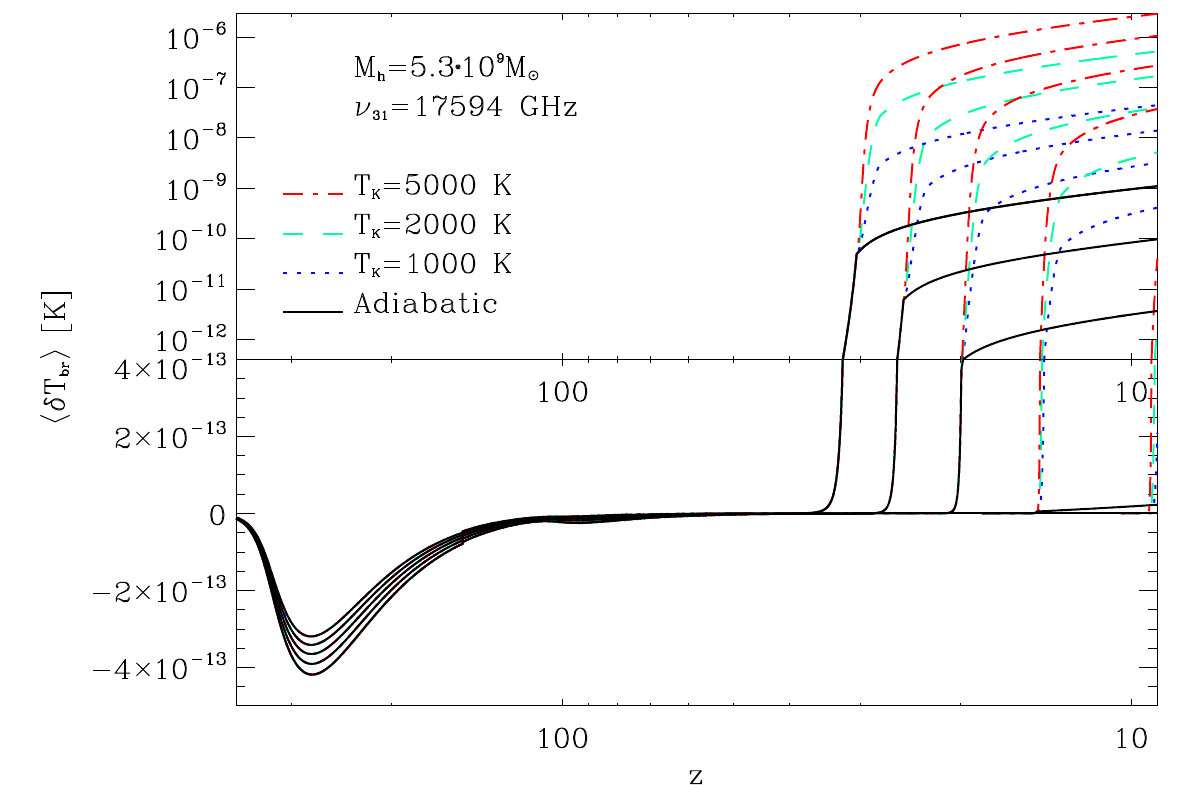}
\includegraphics[width=0.45\textwidth]{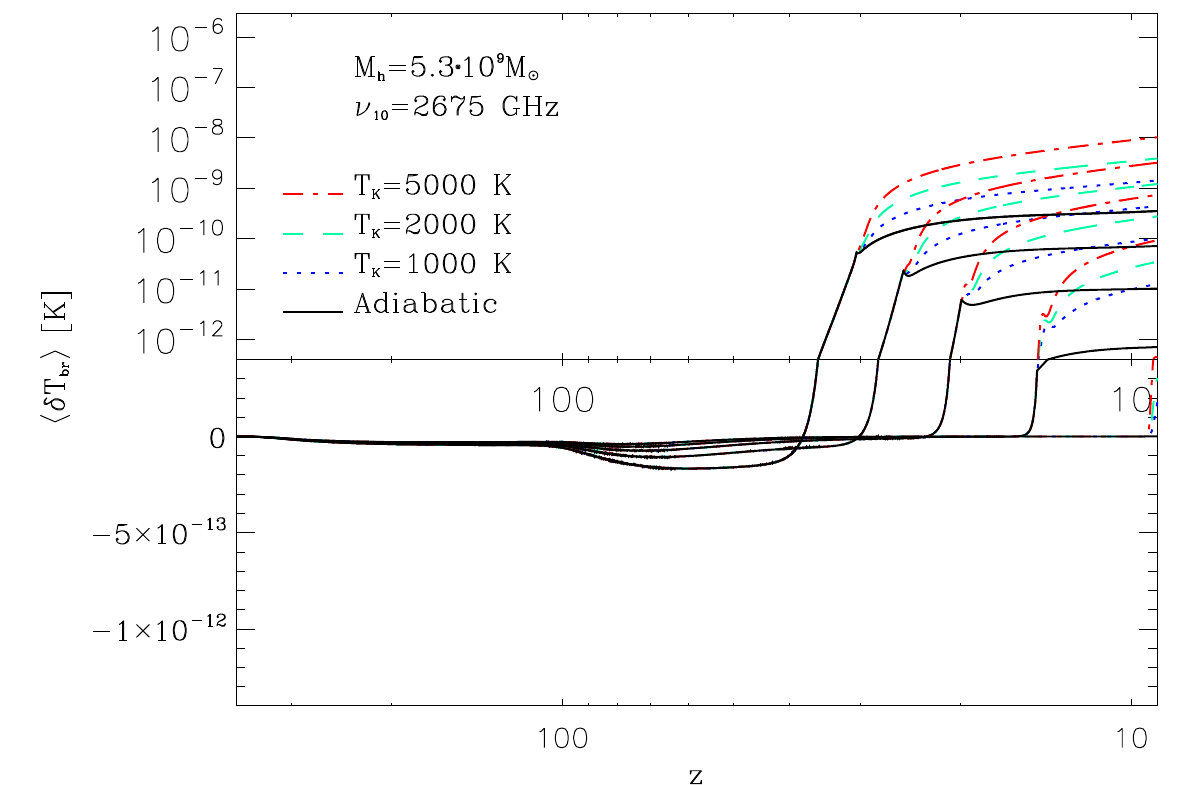}
\caption{Evolution of  brightness temperature in the lines of transitions $J=2\rightarrow0$ and $J=3\rightarrow1$ of molecule H$_2$ and $J=1\rightarrow0$ of molecule HD for halos with mass $M_h=5.3\cdot10^9$ M$_\odot$. Each line corresponds to the halo with different initial amplitude of curvature perturbation like in previous figures and different final temperature. }
\label{Tbe_Tk}
\end{figure}
One can conclude that so small value of expected molecular hydrogen emission from dark ages halos are caused by low values of adiabatic temperature in them. Really, the virialization temperature of halos can be essentially larger than adiabatic temperature caused by smooth homogeneous collapse \citep{Barkana2001,Bromm2011,Novosyadlyj2018}. Local inhomogeneities, shock waves, fragmentation and other nonlinear phenomena will enhance the kinetic temperature of gas in halos. Let's estimate the differential brightness temperature of dark ages halos in the rotational lines of hydrogen molecules for higher kinetic temperatures of gas in the virialized halos which had the same history before. We suppose that halos after virialization reach some temperatures $T_{K(vir)}>T_{K(ad)}$ independent on the redshift of virialization. The results for $T_{K(vir)}$=1000, 2000, 5000 K are presented in Figure \ref{Tbe_Tk}. Really, increasing of kinetic temperature of gas in the virialized halos leads to the increasing of their differential brightness temperature in the rotational lines of molecular hydrogen. We see that for para/ortho-hydrogen molecules the effect of temperature is larger than for hydrogen deuteride ones, and the base line of ortho-H$_2$ is more intensive than para-H$_2$. The values of the differential brightness temperatures of hot  halos ($T_{K(vir)}\gtrsim2000$ K) are comparable with the sensitivity of Planck observatory \citep{Planck2018a} and expected sensitivity of SKA radio telescope that is being built, $\langle\delta T^b_{ul}\rangle\sim10^{-6}$ K. Both, unfortunately, are far from the angular resolution which is necessary to detect individual halos analyzed here.

It is interesting to compare our results with estimations of H$_2$ and HD emission in the same lines from the primordial object, either Pop III galaxies or primordial clouds, which have been computed and discussed by \citet{Kamaya2002,Kamaya2003,Omukai2003,Mizusawa2005}. The expected emission flux from the molecular cloud cores at the redshift of $z\sim$20 in the $J = 2\rightarrow0$ line of H$_2$ is
$\sim2\cdot10^{-7}$ Jy, and in the $J = 4\rightarrow3$ line of HD is $\sim8\cdot10^{-9}$ Jy \citep{Kamaya2003}. The differential spectral fluxes $\delta F$ in these lines at $z\sim$24 from the smallest halos analyzed here are $3\cdot10^{-15}$ and $3\cdot10^{-19}$ Jy accordingly. Since the differential brightness temperature $\delta T_{br}$ in those lines are $1.5\cdot10^{-12}$ K and $1.9\cdot10^{-16}$ K respectively, and $\delta F\propto\delta T_{br}$, and $F\propto T_{br}$ with the same coefficient of proportion, the spectral fluxes in these lines are $\sim5\cdot10^{-3}$ Jy. \citet{Omukai2003} have estimated the energy fluxes in the rotational lines of H$_2$ molecule from forming galaxies at $z=20$: $8\cdot10^{-27}$ erg/cm$^2$s ($J=2\rightarrow0$) and $5.6\cdot10^{-26}$ erg/cm$^2$s ($J=3\rightarrow1$) for galaxies with $M=10^7$ M$_\odot$, and $9\cdot10^{-24}$ erg/cm$^2$s ($J=2\rightarrow0$) and $7.7\cdot10^{-23}$ erg/cm$^2$s ($J=3\rightarrow1$) for galaxies with $M=10^9$ M$_\odot$. According to our results (Table \ref{Tab10} in Appendix), the fluxes in these lines are as follows: $8\cdot10^{-22}$ erg/cm$^2$s ($J=2\rightarrow0$) and $3.6\cdot10^{-21}$ erg/cm$^2$s ($J=3\rightarrow1$) for galaxies with $M=10^7$ M$_\odot$, and $7\cdot10^{-20}$ erg/cm$^2$s ($J=2\rightarrow0$) and $3.2\cdot10^{-19}$ erg/cm$^2$s ($J=3\rightarrow1$) for galaxies with $M=5.3\cdot10^9$ M$_\odot$. So, fluxes in lower rotational lines of hydrogen and hydrogen deuteride molecules from dark ages halos analyzed here are in $\sim$4 order higher than the fluxes in the same lines from other primordial objects. Such large difference is caused mainly by the difference in structural models of sources and the mechanisms of excitation/de-excitation of the rotational levels. In our case the excitation by CMB is dominant, but in foregoing studies the thermal luminescence has been analyzed.    

We estimate also how the predictions of opacities of halos and brightness temperatures in the lines of transitions between rotational levels of ortho-/para-H$_2$ change with collision rate coefficients revised by \cite{Lique2015}. In the line of transition $J=0\rightarrow2$ (para-H$_2$) the opacities increase by 1.2-1.8 times for the halos virialized at $z$=30-10, while in the line of transition $J=1\rightarrow3$ (ortho-H$_2$) the opacities decrease by 1.1-1.7 times. These differences as well as the change the populations  of levels and excitation temperatures lead to change the differential brightness temperature. In the halos with kinetic temperature $\gtrsim170$ K the differential brightness temperature in the base line of para-H$_2$ ($J=2\rightarrow0$) increases by 1.4-1.7 times, for colder halos it increases by 3.5-5.4 times. While, the base line of ortho-H$_2$ ($J=3\rightarrow1$), vise versa, decreases by 1.7-2.6 times for warmer halos and decreases $\sim3.5$ times for colder ones. Hence, the main conclusion of this re-analysis is that revised collisional excitation/de-excitation coefficients do not change the main results and conclusions.

\section{Resonant scattering}

\begin{figure}[!ht]
\includegraphics[width=0.49\textwidth]{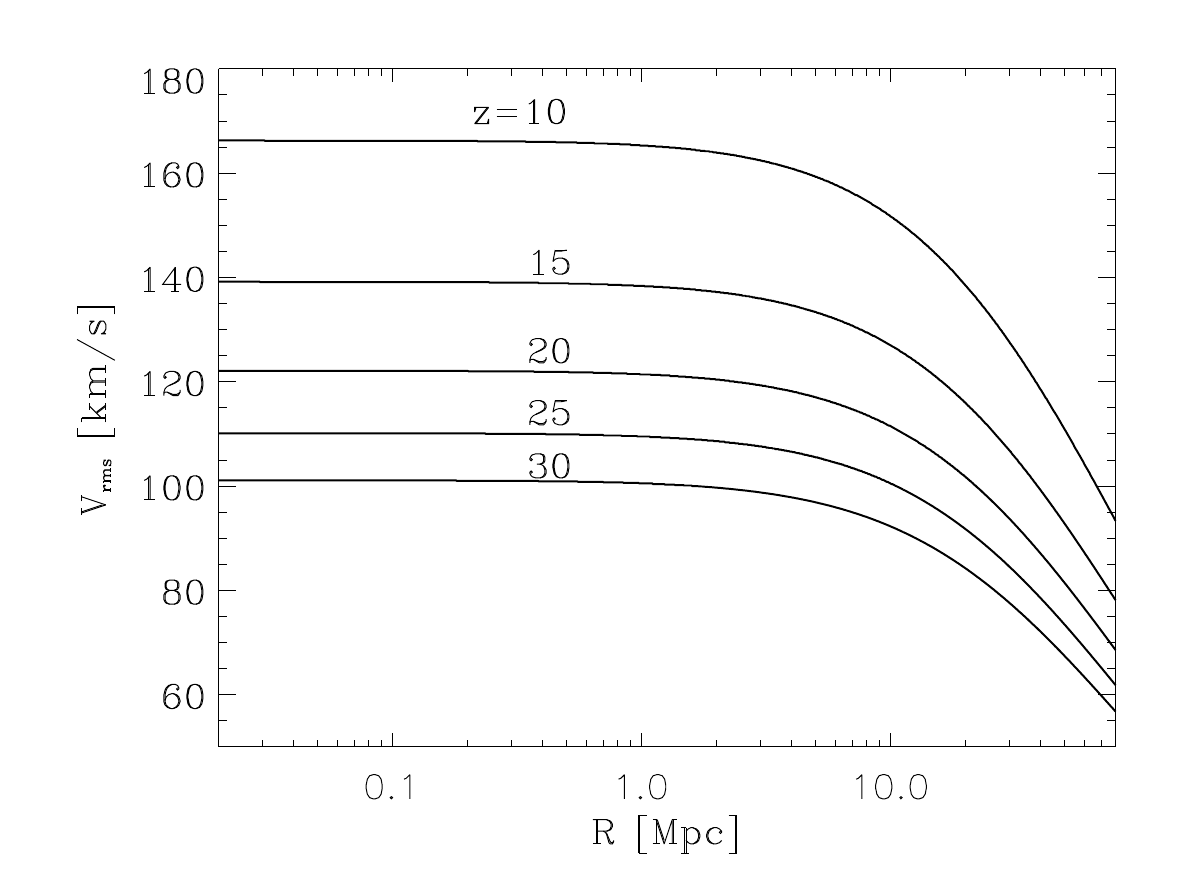}
\caption{The rms peculiar velocity averaged in the top-hat sphere with radius $R$ at different redshifts in the wCDM model with post-Planck cosmological parameters \citep{Planck2018b}.}
\label{Vrms}
\end{figure}

The resonant scattering of CMB quanta in the rotational lines of molecules in the dark ages halo which has optical depth $\tau_{ul}$ and peculiar velocity $v_p$ leads 
to differential brightness temperature \citep{Maoli1996,Persson2010}
\begin{equation}
\delta T^b_{ul}=\frac{h^2\nu_{ul}^2}{k_B^2T_R}\frac{\tau_{ul}}{\left(1-e^{-\frac{h\nu_{ul}}{k_BT_R}}\right)^2}\frac{v_p}{c} \cos\theta, 
\label{Tbr_rs}
\end{equation}
where $\theta$ is the angle between the vector of peculiar velocity and the line of sight of the terrestrial observer. The rms value of peculiar velocity averaged over spherical region with radius $R$ can be estimated if the power spectrum $P(k;z)$ of initial density perturbations is known
\begin{equation}
\langle v^2_p\rangle=\frac{H^2(z)}{2\pi^2(1+z)^2}\int_0^{\infty}P(k;z)W^2(kR)dk,
\label{vp}
\end{equation}
where $W(x)\equiv3(\sin{x}-x\cos{x})/x^3$ is the top-hat sphere in Fourier space. We present the $V_{rms}\equiv \langle v^2_p\rangle^{1/2}$ as function 
of smoothing scale $R$ for power spectrum of wCDM model with post-Planck cosmological parameters \citep{Planck2018a,Planck2018b} in Figure \ref{Vrms}. The power spectrum $P(k;z)\equiv A_s(z)k^{n_s}T^2(k;z)$ we have normalized by computation of $\sigma_8=0.806$ for the current epoch using the fitting formula for the transfer function $T(k;z)$ proposed by \citet{Eisenstein1998}. We rescaled the amplitude $A_s(z)$ for different redshifts using the square of the growth function $D(z)$ by \citet{Carroll1992}.
\begin{table}[ht!]
\begin{center}
\caption{Resonant scattering: the differential brightness temperatures of halos with $M_h=5.3\cdot10^9$ M$_\odot$ virialized at different redshifts $z_v$ in the lowest rotational levels of molecular hydrogen H$2$.}
\begin{tabular} {lcccc}
\hline
\hline
   \noalign{\smallskip}
$z_v$ &\multicolumn{4}{c}{$\langle\delta T^b_{ul}\rangle$ [K]}\\
 \noalign{\smallskip} 
&2-0&3-1&4-2&5-3\\
 \noalign{\smallskip} 
\hline
   \noalign{\smallskip} 
30.41 &1.53$\cdot10^{-11}$&  4.11$\cdot10^{-12}$&  2.13$\cdot10^{-15}$&  4.78$\cdot10^{-18}$ \\
25.15 &3.49$\cdot10^{-12}$&  4.34$\cdot10^{-13}$&  2.92$\cdot10^{-17}$&  1.77$\cdot10^{-20}$ \\
19.90 &4.31$\cdot10^{-13}$&  1.68$\cdot10^{-14}$&  5.57$\cdot10^{-20}$&  5.39$\cdot10^{-24}$ \\
14.65 &1.75$\cdot10^{-14}$&  8.77$\cdot10^{-17}$&  2.09$\cdot10^{-24}$&  1.08$\cdot10^{-29}$ \\
 9.41 &5.15$\cdot10^{-17}$&  2.95$\cdot10^{-21}$&  5.83$\cdot10^{-33}$&  5.36$\cdot10^{-41}$ \\ 
 \noalign{\smallskip} 
  \hline
\end{tabular}
\label{Tab_dTb_H2_rs}
\end{center}
\end{table}
\begin{table}[ht!]
\begin{center}
\caption{Resonant scattering: the differential brightness temperatures of halos with $M_h=5.3\cdot10^9$ M$_\odot$ virialized at different redshifts $z_v$ in the lowest rotational levels of hydrogen deuteride molecules HD.}
\begin{tabular} {ccccc}
\hline
\hline
   \noalign{\smallskip}
$z_v$ &\multicolumn{4}{c}{$\langle\delta T^b_{ul}\rangle$ [K]}\\
 \noalign{\smallskip} 
&1-0&2-1&3-2&4-3 \\
 \noalign{\smallskip} 
\hline
   \noalign{\smallskip}  
30.41 &3.57$\cdot10^{-9}$&  1.17$\cdot10^{-9}$&   4.37$\cdot10^{-11}$&  2.77$\cdot10^{-13}$ \\
25.15 &4.24$\cdot10^{-9}$&  7.98$\cdot10^{-10}$&  1.23$\cdot10^{-11}$&  2.41$\cdot10^{-14}$ \\
19.90 &4.42$\cdot10^{-9}$&  3.56$\cdot10^{-10}$&  1.45$\cdot10^{-12}$&  4.80$\cdot10^{-16}$ \\
14.65 &2.06$\cdot10^{-9}$&  3.86$\cdot10^{-11}$&  1.68$\cdot10^{-14}$&  2.84$\cdot10^{-19}$ \\
 9.41 &3.05$\cdot10^{-10}$& 2.90$\cdot10^{-13}$&  1.40$\cdot10^{-18}$&  6.13$\cdot10^{-26}$ \\ 
 \noalign{\smallskip} 
  \hline
\end{tabular}
\label{Tab_dTb_HD_rs}
\end{center}
\end{table}

We compute the absolute values of differential brightness temperature $\delta T^b_{ul}$ and spectral flux $\delta F_{ul}$ caused by resonant scattering in the same rotational lines of para-H$_2$, ortho-H$_2$ and HD molecules for the same halos. The results are presented in Tables \ref{Tab_dTb_H2_rs}-\ref{Tab_dTb_HD_rs} and Tables \ref{Tab10}-\ref{Tab11} in Appendix of the on-line version of the paper. They show that expected amplitudes of intensities of emission/absorption in the rotational lines of both isomers of molecular hydrogen caused by resonant scattering are by order and more lower than caused by thermal collisions with adiabatically warmed gas in the dark ages halos. And contrary, the amplitudes of halo intensities 
in the rotational lines of HD molecule caused by resonant scattering are systematically higher than thermal emission intensities. The effect is more noticeable at small redshifts ($z\sim10-15$), from where the signal, however, is much weaker. 

Signals caused by resonant scattering increases with opacity of halos like thermal ones, but decrease when radiation and gas temperatures increase. Hence, the warm halos  ($T_K\sim400$ K) are brighter in the HD rotational lines ($\nu_{obs}\sim85$ GHz) caused by resonant scattering, while the hot halos ($T_K\sim2000$ K) are brighter in the line of ortho-H$_2$ ($\nu_{obs}\sim338$ GHz) caused by thermal collisions with hydrogen atoms.

The estimations of the differential brightness temperature caused by the resonant scattering in the first rotational levels of hydrogen deuteride molecule in the dark ages halos presented here are comparable with the values obtained by \citet{Nunez2006}. They computed the secondary anisotropy of CMB caused by resonant scattering in HD lines of top-hat halos of the wide mass range $10^2-10^{12}$ M$_\odot$ at the stages of their turn-around and after virialization. They have constrained ourselves by analysis of the halos which are formed from the overdensity with the initial amplitude of relative density perturbations $\delta\rho/\bar{\rho}\sim3\sigma$ and $6\sigma$, where $\sigma$ is the value of rms density perturbations at the corresponding scale. We study the halos with initial amplitudes of curvature perturbations $C_k=1.5-3\cdot10^{-4}$ (seed of halos) that are in the same range of peak heights (see section 2.1 in \citep{Novosyadlyj2018}). The opacities in the rotational lines $J=1-0$, 2-1 and 3-2 in the left panel of Figure 5 and 9 for halos with mass of $5.3\cdot10^9$ M$_\odot$ and corresponding initial amplitudes are comparable with opacities computed by \citet{Nunez2006} (Figure 2 and 3) for the halos at their turn-around points (in our figures they are at $z=46$, 38, 30, 22 and 15 for lines from top to bottom). The secondary anisotropy of CMB $\Delta T/\bar{T}$ caused by resonant scattering in halos at turn-around point computed by \citet{Nunez2006} is in the range $\sim10^{-10}-10^{-12}$ for the first rotational transition of HD molecule. Our computations give the values $\sim 3\cdot10^{-12}-2\cdot10^{-14}$ for halos of mass $5.3\cdot10^9$ M$_\odot$ and initial amplitudes from the range 3-6$\sigma$. They will be larger for the halos of larger masses. For the virialized halos \citet{Nunez2006} obtained $\sim 2\cdot10^{-5}-10^{-8}$. Our estimations for hot halos fall within this range of values (Figure \ref{Tbe_Tk}).

Since the peculiar velocities of different halos have a random amplitude and direction, the superposition of these effects gives the rms total amplitude of the differential brightness temperature 
$\langle\delta T^b_{ul}\rangle=\left[\langle\left(\delta T^{b(th)}_{ul}\right)^2\rangle+\langle\left(\delta T^{b(rs)}_{ul}\right)^2\rangle\right]^{1/2}$, where angle parenthesis means ensemble averaging.

\section{Conclusions}

We have analyzed the emission of dark ages halos in the lines of transitions between lowest rotational levels of hydrogen molecules, para-H$_2$, ortho-H$_2$ and HD. It was assumed that halos are homogeneous top-hat  spheres formed from the cosmological density perturbations in the four-component Universe with post-Planck cosmological parameters. We have considered the excitation/de-excitation of lowest five rotational levels of hydrogen molecules by cosmic microwave background radiation and thermal collisions with atomic hydrogen. The kinetic equations for populations of the energy levels have been solved by two independent ways, which give the close final values of differential brightness temperatures of virialized halos caused by thermal collisions. The first one consists in the integration of system of differential kinetic equations for populations of the levels together with system of differential equations of evolution of cosmological perturbations and kinetic equations of formation/destruction of the first molecules. The results of numeric integration, presented in Figures \ref{Tbe_H2}-\ref{Tbe_Tk} and 8-9 in Appendix of the on-line version, show the dependence of opacities and differential brightness temperatures on the redshift from the early stage of halo formation to virialized final stage. The second way of analysis consist in solution of system of algebraic equations for populations  of the lowest rotational levels of molecules H$_2$ and HD in the virialized halos. These solutions give us a possibility to compute the excitation temperatures of levels, the differential brightness temperatures and fluxes in the lines of transitions between them, which are presented in Tables \ref{Tab3}-\ref{Tab_dTb_HD} and \ref{Tab10}-\ref{Tab11} in Appendix of the on-line version. Both ways give practically the same values for populations of levels, opacities, differential brightness temperatures and spectral fluxes for virialized halos. We also compute the differential brightness temperatures and spectral fluxes in the same lines caused by resonant scattering of cosmic microwave background radiation for the halos with a root mean square value of peculiar velocity at different redshift. The results are presented in Tables \ref{Tab_dTb_H2_rs}-\ref{Tab_dTb_HD_rs} and in the last two columns of Tables \ref{Tab10}-\ref{Tab11} in Appendix of the on-line version.  

The main conclusions issued from the obtained results are as follows. 1) The dark ages halos are source of emission in the lines of transitions between lowest rotational levels of hydrogen molecules excited by CMB radiation and collisions with atoms of neutral hydrogen. 2) The amplitudes of differential brightness temperatures caused by thermal collisions with H are systematically larger in the rotational lines of ortho-H$_2$ and para-H$_2$ molecules than in the lines of HD ones. The maximal values are reached for the earliest warm small halos ($z_v\sim50$, $T_K\sim800$ K, $M_h\sim10^6$ M$_{\odot}$)  but they do not exceed nano Kelvin at $\nu_{obs}\sim200-300$ GHz. 3) The amplitudes of differential brightness temperatures in the lowest rotational lines of HD molecules caused by resonant scattering are systematically higher than caused by thermal collisions with H for halos with peculiar velocity equal to rms one followed from the power spectrum of cosmological perturbations and directed to/from the terrestrial observer. The  maximal values which are reached are about few nanokelvins at $\nu_{obs}\sim85-170$ GHz for the massive warm halos ($z_v\sim20$, $T_K\sim200$ K, $M_h\sim5\cdot10^9$ M$_{\odot}$).  4) If the dark ages halos are hot after virialization ($T_K\sim 2000-5000$) then the differential brightness temperatures caused by thermal collisions are increased by a few orders large, reaching the values in few microkelvins, that may be achievable for the next-generation telescopes.  

\acknowledgments
This work was supported by the International Center of Future Science and College of Physics of Jilin University (P.R.China) and the project of Ministry of Education and Science of Ukraine ``Formation and characteristics of elements of the structure of the multicomponent Universe, gamma radiation of supernova remnants and observations of variable stars'' (state registration number 0119U001544). We acknowledge the anonymous referee for his accurate report and useful comments and suggestions.

\newpage

\section*{Appendix A. Tables and figures for the online version only.}

\begin{table*}[ht!]
\caption{Physical values and chemical composition of halos virialized at different $z_v$: $M$ is the total mass, $C_k$ is the amplitude of initial curvature perturbation (seed of halo), $z_v$ is the redshift of virialization, $\rho_{m}$ is the matter density virialized halo, $T_K$ is kinetic temperature of baryonic gas, $n_{HI}$ is the number density of neutral hydrogen atoms, $n_p,\,n_e$ are the number densities of protons and electrons at $z=z_v/10$, $n_{H_2}$ and $n_{HD}$ are the number densities of molecules H$_2$ and HD, $r_h$ is the radius of halo in comoving coordinates, $\theta_h$ is the angular radius of geometrically limited halo.}
\begin{tabular} {cccccccccccc}
\hline
\hline
   \noalign{\smallskip}
$M_h$&$k$&$C_k$&$z_v$&$\rho_{m}$&$T_K$&$n_{HI}$&$n_p\approx n_e$&$n_{H_2}$&$n_{H_D}$&$r_h$&$\theta_h$\\
 \noalign{\smallskip} 
[M$_{\odot}$]&[Mpc$^{-1}$]& & &[g/cm$^3$]&[K]&[cm$^{-3}$]&[10$^{-6}$cm$^{-3}$]&[10$^{-6}$cm$^{-3}$]&[10$^{-9}$cm$^{-3}$]&[kpc]&['']\\
 \noalign{\smallskip} 
\hline
   \noalign{\smallskip}
$6.6\cdot10^8$&10&$3.0\cdot10^{-4}$&35.54&$2.40\cdot10^{-23}$&508.9&1.80&156.3/3.7&30.32&3.83 &0.77&0.50\\ 
 \noalign{\smallskip}
              &  &$2.5\cdot10^{-4}$&29.41&$1.38\cdot10^{-23}$&382.8&1.04& 97.4/3.8&12.41&2.36 &0.92&0.52\\ 
 \noalign{\smallskip} 
              &  &$2.0\cdot10^{-4}$&23.28&$7.05\cdot10^{-24}$&266.3&0.53& 54.2/3.9&4.29&1.94 &1.15&0.53\\ 
 \noalign{\smallskip}
              &  &$1.5\cdot10^{-4}$&17.17&$2.95\cdot10^{-24}$&162.1&0.22& 25.1/4.2&1.18&0.98 &1.54&0.56\\ 
 \noalign{\smallskip}
              &  &$1.0\cdot10^{-4}$&11.05&$8.60\cdot10^{-25}$& 78.2&0.06&  8.3/4.7&0.23&0.16 &2.32&0.61\\             
 \noalign{\smallskip}
 \noalign{\smallskip} 
 $8.3\cdot10^7$&20&$3.0\cdot10^{-4}$&40.57&$3.54\cdot10^{-23}$&619.2&2.65&214.5/3.7&58.54&6.24 &0.34&0.25\\ 
 \noalign{\smallskip}
              &  &$2.5\cdot10^{-4}$&33.55&$2.03\cdot10^{-23}$&468.4&1.52&133.5/3.7&23.71&3.25 &0.40&0.25\\
 \noalign{\smallskip}
              &  &$2.0\cdot10^{-4}$&26.54&$1.03\cdot10^{-23}$&325.5&0.77& 74.0/3.8&7.96&2.10 &0.51&0.26\\    
  \noalign{\smallskip}
              &  &$1.5\cdot10^{-4}$&19.50&$4.24\cdot10^{-24}$&199.6&0.32& 33.8/4.0&2.08&1.46 &0.68&0.27\\ 
   \noalign{\smallskip}
              &  &$1.0\cdot10^{-4}$&12.36&$1.17\cdot10^{-24}$& 94.0&0.09& 10.4/4.5&0.36&0.30 &1.05&0.30\\    
 \noalign{\smallskip}
 \noalign{\smallskip}             
$1\cdot10^7$&40&$3.0\cdot10^{-4}$ &45.72&$5.01\cdot10^{-23}$&730.3&3.76&287.2/3.6&103.8&9.94 &0.15&0.12\\ 
    \noalign{\smallskip}
              & &$2.5\cdot10^{-4}$&37.85&$2.88\cdot10^{-23}$&556.0&2.16&179.7/3.7&40.91&4.78 &0.18&0.13\\ 
    \noalign{\smallskip}
              & &$2.0\cdot10^{-4}$&29.92&$1.46\cdot10^{-23}$&392.9&1.09& 99.7/3.8&13.97&2.45 &0.23&0.13\\  
    \noalign{\smallskip}
              & &$1.5\cdot10^{-4}$&22.02&$6.00\cdot10^{-24}$&242.8&0.45& 45.7/3.9&3.53&1.80 &0.30&0.13\\  
    \noalign{\smallskip}
              & &$1.0\cdot10^{-4}$&14.00&$1.66\cdot10^{-24}$&115.2&0.13&14.1/4.3 &0.58&0.50 &0.47&0.14\\ 
 \noalign{\smallskip}
 \noalign{\smallskip}
$1.3\cdot10^6$&80&$3.0\cdot10^{-4}$&50.33&$6.66\cdot10^{-23}$&834.0&5.00&355.3/3.6&172.5&15.40 &0.068&0.061\\ 
 \noalign{\smallskip} 
              &  &$2.5\cdot10^{-4}$&41.52&$3.78\cdot10^{-23}$&636.5&2.84&218.8/3.7&69.78&7.28 &0.082&0.062\\ 
 \noalign{\smallskip} 
              &  &$2.0\cdot10^{-4}$&32.65&$1.87\cdot10^{-23}$&446.8&1.41&117.7/3.7&23.12&3.25 &0.10 &0.064\\ 
 \noalign{\smallskip}
              &  &$1.5\cdot10^{-4}$&24.40&$8.07\cdot10^{-24}$&286.1&0.61& 58.6/3.9&5.68&1.94 &0.14 &0.066\\ 
 \noalign{\smallskip}
              &  &$1.0\cdot10^{-4}$&15.38&$2.16\cdot10^{-24}$&134.4&0.16& 17.1/4.2&0.89&0.78 &0.21&0.071\\              
  \hline
\end{tabular}
\label{Tab1A}
\end{table*}

\begin{table*}[ht!]
\caption{The best-fit coefficients of analytical approximation of the dependences of the collisional deactivation rate coefficients on the kinetic temperature $T_K$ by formula 
$\lg{\kappa_{ul}(T_K)}=\sum_{k=0}^{N-1}a^{(ul)}_k(\lg{T_K/\epsilon^{(ul)}})^k+a^{(ul)}_N(1/(T_K/\epsilon^{(ul)}+\epsilon^{(ul)})-1$  for H$_2$ molecule (solid lines in the left panel of Figure \ref{VAMDCkappa}). It is valid in the temperature range $100\le T_K\le6000$ K. The coefficients are taken from the VAMDC basedata, the misprints in the formula have been fixed. }
\begin{tabular} {cccccccccccc}
\hline
\hline
   \noalign{\smallskip}
$u$&$l$&$\epsilon$&$a_0$&$a_1$&$a2$&$a_3$&$a_4$&$a_5$&$a_6$&$a_7$&$a_8$\\
 \noalign{\smallskip}
\hline
   \noalign{\smallskip}    
 2& 0&  6.0& -8259.36044&  1846.12284& -4713.51503&  3907.15150& -1672.56520&  4015.80145& -51.6262545&  2.78079268& -9095.89801\\
 3& 1& 15.0&  1402.43596& -58.1010752&  74.2671058&  18.8773370& -36.2141786&  12.2012441& -1.32412404&  0.00000000&  1484.70110\\
 4& 2& 15.0&  11533.9595&  2070.41178& -4153.50389&  5002.05894& -3171.93991&  1091.80410& -195.457394&  14.3495151&  12713.2249\\
 5& 3& 13.0&  5064.45306&  73.5896891& -266.094034&  722.660337& -590.299001&  223.156022& -41.0457115&  2.98641025&  5439.61121\\
 4& 0& 24.0&  6887.91049& -219.803978&  673.503903& -912.827587&  802.461038& -395.623746&  98.4201969& -9.67987032&  7148.50271\\
 5& 1& 61.0&  76184.3307&  76.4032424& -87.2258173&  387.822432& -476.720251&  397.196415& -157.426338&  22.3504536&  77451.1227\\
  \noalign{\smallskip}
  \hline
\end{tabular}
\label{TabH2deexcfit}
\end{table*}

\begin{table}[ht!]
\begin{center}
\caption{The best-fit coefficients of analytical approximation of the dependences of the collisional deactivation rate coefficients on the kinetic temperature $T_K$ by quadratic parabola for H$_2$ molecule for temperature range 50 -500 K (solid lines in the left panel of Figure \ref{VAMDCkappa}).}
\begin{tabular} {ccccc}
\hline
\hline
   \noalign{\smallskip}
$u$&$l$&$a_0$&$a_1$&$a2$ \\
 \noalign{\smallskip}
\hline
   \noalign{\smallskip}    
 2& 0&-14.653&0.066487&0.24931 \\
 3& 1&-6.0491&-6.8089&1.6397 \\
 4& 2&-9.5234&-3.7050&0.94767 \\
 5& 3&-13.942&-1.0658&0.56254 \\
 4& 0&-14.910&0.072763&0.14760 \\
 5& 1&-18.092&0.1.4405&0.063031 \\
  \noalign{\smallskip}
  \hline
\end{tabular}
\label{TabH2lowTcdexc}
\end{center}
\end{table}

\begin{table}[ht!]
\begin{center}
\caption{The best-fit coefficients of analytical approximation of the dependences of the collisional deactivation rate coefficients for HD molecule on the kinetic temperature $T_K$ by polynom of 3d order, $\lg{\kappa_{ul}(T_K)}=a_{ul}+b_{ul}\lg{T_K}+c_{ul}(\lg{T_K})^2+d_{ul}(\lg{T_K})^3$ (solid lines in the right panel of Figure \ref{VAMDCkappa}).}
\begin{tabular} {cccccc}
\hline
\hline
   \noalign{\smallskip}
$u$&$l$&$a$&$b$&$c$&$d$\\
 \noalign{\smallskip}
\hline
   \noalign{\smallskip} 
 1& 0& -16.026& -2.2319&  1.2578& -0.17135\\
 2& 0& -12.813& -8.2155&  3.6827& -0.46921\\
 3& 0& -6.1385& -16.448&  6.3996& -0.74128\\ 
 4& 0&  8.5527& -32.399&  11.634& -1.2838\\
 5& 0&  2.6510& -27.177&  9.7365& -1.0395\\ 
 2& 1& -15.293& -2.7751&  1.3979& -0.18162\\
 3& 1& -12.970& -7.7509&  3.4431& -0.43110\\
 4& 1& -5.8105& -16.797&  6.5236& -0.75144\\
 5& 1& -2.9747& -21.584&  8.2829& -0.93604\\ 
 3& 2& -13.998& -4.3688&  2.0205& -0.25961\\
 4& 2& -13.766& -6.9609&  3.1745& -0.39987\\
 5& 2& -9.2775& -13.887&  5.7245& -0.67937\\
 4& 3& -16.229& -2.2922&  1.3529& -0.18594\\
 5& 3& -14.639& -6.3194&  3.0045& -0.38352\\
 5& 4& -16.554& -2.3252&  1.4452& -0.20237\\
\noalign{\smallskip}
  \hline
\end{tabular}
\label{TabHDTcdexc}
\end{center}
\end{table}

\animategraphics[width=\textwidth]{1}{figure_H2}{0}{4}

\animategraphics[width=\textwidth]{1}{figure_HD}{0}{4}

\begin{table*}[!ht]
\begin{center}
\caption{The optical depths, brightness temperatures and spectral fluxes in the two lowest rotational lines of molecule H$_2$ in the dark ages halos of different masses $M_h$ virialized at different redshift $z_v$. Marking (th) means the thermal emission, marking (rs) means the resonant scattering.} 
\begin{tabular} {ccccccccc}
\hline
\hline
   \noalign{\smallskip}
$M_h$&$z_v$&$\nu_{obs}$&$\Delta\nu_{obs}$&$\tau_{\nu}$&$\delta T_{br}^{(th)}$&$\delta F^{(th)} $&$\delta T_{br}^{(rs)}$&$\delta F^{(rs)} $ \\
 \noalign{\smallskip} 
[M$_{\odot}$]& &[GHz]&[kHz]&$10^{-9}$ &[$10^{-12}$K] &[$10^{-12}$Jy]&[$10^{-12}$K] &[$10^{-12}$Jy] \\
 \noalign{\smallskip} 
\hline
   \noalign{\smallskip} 
1.29$\cdot10^6$&   50.33&  207&  3.02&1.88&  799.0&  0.288&  16.9&  0.006 \\
           &        &  343&  4.99&13.1&  657.0&  0.650&  28.1&  0.028 \\
           &   41.52&  250&  3.18&1.39&  175.0&  0.095&  7.55&  0.004 \\
           &        &  414&  5.27&7.44&   91.1&  0.136&  6.05&  0.009 \\
           &   32.65&  316&  3.37&0.77&   22.0&  0.020&  1.82&  0.002 \\
           &        &  523&  5.58&3.76&   7.49&  0.019&  0.63&  0.002 \\
           &   24.40&  418&  3.57&0.31&   1.51&  0.003&  0.18&  0.0003 \\
           &        &  693&  5.91&1.55&   0.30&  0.001&  0.02&  0.00009 \\
           &   15.39&  648&  3.79&0.11&   0.01&  0.00005&  0.002& 0.00001 \\
           &        & 1070&  6.28&0.54&   0.0008&0.00001&  0.00002&0.0000002 \\ 
\noalign{\smallskip}                     
1.04$\cdot10^7$&   45.72&  227&  3.10&3.16&  732.0&  1.300&  22.4&  0.040 \\
           &        &  377&  5.14&18.7&  468.0&  2.280&  25.4&  0.124 \\
           &   37.85&  273&  3.25&2.03&  141.0&  0.380&  8.29&  0.022 \\
           &        &  453&  5.39&10.3&   61.9&  0.458&  4.82&  0.036 \\
           &   29.92&  343&  3.44&1.08&   18.6&  0.083&  1.78&  0.008 \\
           &        &  569&  5.69&5.29&   5.45&  0.066&  0.45&  0.005 \\
           &   22.02&  461&  3.63&0.46&   1.09&  0.009&  0.15&  0.001 \\
           &        &  764&  6.01&2.31&   0.18&  0.004&  0.01&  0.00023 \\
           &   14.00&  708&  3.83&0.18&  0.006&  0.0001&  0.001& 0.00003 \\
           &        & 1170&  6.35&0.82&  0.0004& 0.00002& 0.000005&0.0000003 \\
\noalign{\smallskip}                     
8.28$\cdot10^7$&   40.57&  255&  3.21&4.91&  545.0&  4.990&  24.9& 0.229 \\
           &        &  423&  5.31&25.9&  273.0&  6.870&  18.4& 0.464 \\
           &   33.55&  307&  3.36&2.98&  103.0&  1.420&  7.88& 0.109 \\
           &        &  509&  5.56&14.6&   36.8&  1.400&  3.02& 0.115 \\
           &   26.54&  386&  3.51&1.52&   12.5&  0.288&  1.41& 0.033 \\
           &        &  639&  5.82&7.48&   2.91&  0.184&  0.22& 0.014 \\
           &   19.50&  518&  3.69&0.67&   0.61&  0.028&  0.093&0.004 \\
           &        &  858&  6.12&3.38&   0.078& 0.010&  0.003&0.0004 \\
           &   12.36&  795&  3.89&0.32&  0.0029& 0.0004&  0.0006&0.00008 \\
           &        & 1320&  6.44&1.24&  0.0001& 0.00003&  0.0000007&0.0000002 \\
\noalign{\smallskip}                     
6.63$\cdot10^8$&   35.54&  291&  3.31&6.82&  330.0&  16.1&  22.4& 1.09 \\
           &        &  481&  5.48&33.9&  131.0&  17.5&  10.5& 1.40 \\
           &   29.41&  349&  3.45&3.96&   61.5&  4.53&  6.04& 0.445 \\
           &        &  579&  5.71&19.4&   17.5&  3.530&  1.42&0.288 \\
           &   23.28&  437&  3.60&2.03&   7.27&  0.893&  0.91&0.112 \\
           &        &  725&  5.97&10.2&   1.32&  0.445&  0.080&0.027 \\
           &   17.17&  585&  3.76&0.96&   0.27&  0.065&  0.050&0.012 \\
           &        &  968&  6.22&4.78&   0.027& 0.018&  0.0008& 0.00052 \\
           &   11.05&  881&  3.93&0.52&  0.0011& 0.0007&  0.0003& 0.00017 \\
           &        & 1460&  6.52&1.84&  0.00002&0.00004&  0.0000001& 0.0000002 \\
\noalign{\smallskip}                     
5.30$\cdot10^9$&   30.41&  338&  3.42&8.67&  164.0&  44.9&  15.3& 4.20 \\
           &        &  560&  5.67&42.4&   49.2&  37.0&  4.11& 3.09 \\
           &   25.15&  406&  3.54&4.99&   29.2&  12.1&  3.49& 1.45 \\
           &        &  673&  5.87&24.8&   6.13&   6.98&  0.43&0.494 \\ 
           &   19.90&  508&  3.68&2.67&   2.87&   2.00&  0.43&0.299 \\
           &        &  842&  6.10&13.5&  0.38&    0.729&  0.017&0.032 \\ 
           &   14.65&  679&  3.82&1.43&  0.076&   0.104&  0.018&0.024 \\
           &        & 1120&  6.32&6.64&  0.0051&  0.019&  0.00009&0.0003 \\
           &    9.41& 1020&  3.98&0.94&  0.00016& 0.0006&  0.00005&0.0002 \\
           &        & 1690&  6.60&2.60&  0.000001&0.00001&  0.000000003&0.00000003 \\
  \hline
\end{tabular}
\label{Tab10}
\end{center}
\end{table*}

\begin{table*}[!ht]
\begin{center}
\caption{The optical depths, brightness temperatures and spectral fluxes in the two lowest rotational lines of molecule HD in the dark ages halos of different masses $M_h$ virialized at different redshift $z_v$. Marking (th) means the thermal emission, marking (rs) means the resonant scattering.} 
\begin{tabular} {ccccccccc}
\hline
\hline
   \noalign{\smallskip}
$M_h$&$z_v$&$\nu_{obs}$&$\Delta\nu_{obs}$&$\tau_{\nu}$&$\delta T_{br}^{(th)}$&$\delta F^{(th)} $&$\delta T_{br}^{(rs)}$&$\delta F^{(rs)} $ \\
 \noalign{\smallskip} 
[M$_{\odot}$]& &[GHz]&[kHz]&$10^{-9}$  &[$10^{-12}$K] &[$10^{-12}$Jy]&[$10^{-12}$K] &[$10^{-12}$Jy] \\
 \noalign{\smallskip} 
\hline
   \noalign{\smallskip} 
1.29$\cdot10^6$&   50.33&  52.1&  0.62&  19.0&  230.0&  0.005&  435.0& 0.010\\ 
           &        &   104&  1.23&  21.4&   65.4&  0.006&  401.0& 0.036\\ 
           &   41.52&  62.9&  0.65&  16.3&   99.2&  0.003&  328.0& 0.011\\ 
           &        &   125&  1.30&  14.4&   21.5&  0.003&  217.0& 0.030\\ 
           &   32.65&  79.5&  0.69&  14.8&   37.0&  0.002&  250.0& 0.015\\ 
           &        &   158&  1.38&  9.13&   5.30&  0.001&   98.2& 0.023\\ 
           &   24.40&   105&  0.73&  20.0&   16.0&  0.002&  260.0& 0.029\\ 
           &        &   210&  1.46&  7.25&   1.20&  0.0005&   44.3& 0.019\\   
           &   15.39&   163&  0.78&  25.9&   2.69&  0.0008&  186.0& 0.057\\ 
           &        &   325&  1.55&  3.09&  0.044&  0.00005&  4.53&  0.006\\  
\noalign{\smallskip}                     
1.04$\cdot10^7$&   45.72&  57.3&  0.64&  33.1&  285.0&  0.032&  713.0& 0.08\\ 
           &        &   114&  1.27&  33.2&   71.4&  0.032&  561.0& 0.25\\ 
           &   37.85&  68.9&  0.67&  28.4&  121.0&  0.021&  537.0& 0.92\\ 
           &        &   137&  1.33&  22.0&   22.7&  0.015&  295.0& 0.20\\   
           &   29.92&  86.5&  0.71&  28.7&   51.6&  0.015&  451.0& 0.13\\ 
           &        &   172&  1.41&  15.3&   6.20&  0.007&  141.0& 0.16\\  
           &   22.02&   116&  0.74&  49.0&   25.6&  0.014&  570.0& 0.31\\ 
           &        &   232&  1.48&  14.3&   1.45&  0.032&   67.9& 0.15\\   
           &   14.00&   178&  0.79&  41.6&   2.75&  0.041&  253.0& 0.38\\ 
           &        &   355&  1.57&  3.75&   0.03&  0.0002&   3.68& 0.022\\ 
\noalign{\smallskip}                     
8.28$\cdot10^7$&   40.57&  64.3&  0.66&  59.7&  337.0&  0.196&  1190.0&  0.69\\ 
           &        &   128&  1.31&  51.0&   70.7&  0.163&   749.0& 1.73\\ 
           &   33.55&  77.4&  0.69&  54.8&  153.0&  0.134&   944.0& 0.83\\ 
           &        &   154&  1.37&  35.2&   23.1&  0.081&   396.0& 1.38\\ 
           &   26.54&  97.1&  0.72&  69.1&   77.0&  0.113&   973.0& 1.42\\ 
           &        &   194&  1.44&  29.5&   7.08&  0.041&   217.0& 1.26\\ 
           &   19.50&   130&  0.76& 109.0&   33.9&  0.097&  1100.0& 3.17\\ 
           &        &   260&  1.51&  24.1&   1.33&  0.015&    81.3& 0.93\\ 
           &   12.36&   200&  0.80&  63.8&   2.24&  0.018&   304.0& 2.43\\ 
           &        &   399&  1.59&  3.86&  0.013&   0.0004&    2.08& 0.66\\  
\noalign{\smallskip}                     
6.63$\cdot10^8$&   35.54&  73.2&  0.68& 109.0&  376.0&  1.160&  1960.0& 6.08\\ 
           &        &   146&  1.35&  76.7&   63.1&  0.777&   940.0&11.6\\ 
           &   29.41&  88.0&  0.71& 116.0&  195.0&  0.913&  1800.0& 8.41\\ 
           &        &   175&  1.41&  60.0&   22.7&  0.421&   539.0&10.0\\   
           &   23.28&   110&  0.74& 182.0&  120.0&  0.939&  2250.0&17.5\\ 
           &        &   220&  1.47&  59.8&   7.98&  0.247&   327.0&10.1\\   
           &   17.17&   147&  0.77& 201.0&   35.4&  0.541&  1700.0&26.0\\ 
           &        &   293&  1.54&  32.3&   0.90&  0.054&    71.6& 4.35\\ 
           &   11.05&   222&  0.81&  88.5&   1.68&  0.069&   328.0&13.5\\ 
           &        &   442&  1.61&  3.62&  0.0053& 0.0009&    1.06& 0.17\\ 
\noalign{\smallskip}                     
5.30$\cdot10^9$&   30.41&  85.2&  0.70& 225.0&  428.0& 7.44&  3570.0& 62.1\\ 
           &        &   170&  1.40& 123.0&   53.2& 3.68&  1170.0& 80.9\\ 
           &   25.15&   102&  0.73& 317.0&  284.0& 7.48&  4240.0&112.0\\ 
           &        &   204&  1.45& 122.0&   22.9& 2.40&   798.0& 83.5\\ 
           &   19.90&   128&  0.76& 428.0&  145.0& 6.41&  4420.0&195.0\\ 
           &        &   255&  1.51&  99.3&   60.8& 1.07&   356.0& 62.3\\   
           &   14.65&   171&  0.78& 312.0&   25.9& 2.25&  2060.0&179.0\\ 
           &        &   341&  1.56&  32.3&   0.34& 0.12&    38.6& 13.4\\ 
           &    9.41&   257&  0.82& 125.0&   0.90& 0.21&   305.0& 72.3\\ 
           &        &   512&  1.63&  2.74&   0.001&0.0009&    0.29& 0.27\\
  \hline
\end{tabular}
\label{Tab11}
\end{center}
\end{table*}  

\end{document}